\documentclass[conference]{IEEEtran}
\ifCLASSINFOpdf
\else
\fi
\usepackage{tikz}
\usepackage{multirow}
\usepackage{multicol}
\usepackage{amsmath}
\usepackage{nicefrac}
\usepackage{xspace}
\usepackage{caption}
\usepackage{subcaption}
\usepackage{placeins}

\usepackage{algorithm}
\usepackage{algpseudocode}
\usepackage{numprint}
\usepackage{graphicx}
\usepackage{xcolor}
\usepackage{tcolorbox}
\usepackage{tabularx}
\usepackage{array}
\usepackage{colortbl}

\usepackage{soul}
\usepackage{enumitem}
\tcbuselibrary{skins}
\usepackage{hyperref}

\usepackage{makecell}

\newcommand{\ournameNoSpace}{\mbox{SafeSplit}}
\newcommand{\ournameGen}{\ournameNoSpace's\xspace}
\newcommand{\ourname}{\ournameNoSpace\xspace}

\newcommand{\paperTitle}{\ournameNoSpace: A Novel Defense Against Client-Side Backdoor Attacks in Split Learning}

\newcommand{\adversaryNoSpace}{\ensuremath{\mathcal{A}}}

\newcommand{\adversary}{\adversaryNoSpace\xspace}

\newcommand{\serverNoSpace}{\ensuremath{\mathcal{S}}}
\newcommand{\server}{\serverNoSpace\xspace}

\newcommand{\RDM}{\mbox{$R_{D}$}\xspace}

\newcommand{\etal}{\emph{et~al.}\xspace}
\newcommand{\sect}{Sect.~}

\newcommand{\nonIidNoSpace}{non-IID}
\newcommand{\nonIid}{\nonIidNoSpace\xspace}

\newcommand{\iidNoSpace}{IID}
\newcommand{\iid}{\iidNoSpace\xspace}
\newcommand{\lnorm}{\mbox{\ensuremath{L_2-\text{norm}}}\xspace}

\newcommand{\sota}{\mbox{state-of-the-art} }

\newcommand{\challenge}[1]{C#1}

\newcommand{\nClients}{\ensuremath{N}\xspace}

\newcommand{\scaleTable}[1]{\scalebox{0.9}{#1}}

\newcommand{\cifar}{\mbox{CIFAR-10}\xspace}
\newcommand{\mnist}{MNIST\xspace}
\newcommand{\fmnist}{FMNIST\xspace}
\newcommand{\cifarHundred}{CIFAR-100\xspace}
\newcommand{\gtsrb}{GTSRB\xspace}

\newcommand{\STAB}[1]{\begin{tabular}{@{}c@{}}#1\end{tabular}}

\begin{document}
%

\title{\paperTitle\xspace (Full Version)*}
\author{
\centering\IEEEauthorblockN{Phillip Rieger}
\IEEEauthorblockA{\hspace{-0.5cm}Technical University of Darmstadt}
\and
\IEEEauthorblockN{Alessandro Pegoraro}
\IEEEauthorblockA{\hspace{-0.cm}Technical University of Darmstadt}
\and
\IEEEauthorblockN{Kavita Kumari\hspace{0.5cm}}
\IEEEauthorblockA{\hspace{-0.cm}Technical University of Darmstadt\hspace{.8cm}}\\
\and
\IEEEauthorblockN{\hspace{.6cm}Tigist Abera}
\IEEEauthorblockA{\hspace{.6cm}Technical University of Darmstadt}
\and
\IEEEauthorblockN{\hspace{.15cm}Jonathan Knauer}
\IEEEauthorblockA{\hspace{.15cm}Technical University of Darmstadt}
\and
\IEEEauthorblockN{\hspace{.5cm}Ahmad-Reza Sadeghi}
\IEEEauthorblockA{\hspace{.5cm}Technical University of Darmstadt}
}
	

%


\IEEEoverridecommandlockouts
\makeatletter\def\@IEEEpubidpullup{6.5\baselineskip}\makeatother
\IEEEpubid{\parbox{\columnwidth}{
		Network and Distributed System Security (NDSS) Symposium 2025\\
		24-28 February 2025, San Diego, CA, USA\\
		ISBN 979-8-9894372-8-3\\
		https://dx.doi.org/10.14722/ndss.2025.241698\\
		www.ndss-symposium.org
}
\hspace{\columnsep}\makebox[\columnwidth]{}}

\maketitle
\renewcommand{\thefootnote}{*}

\footnotetext[1]{Please cite the version of this paper published at NDSS 2025~\cite{rieger25safesplit}.}

\renewcommand{\thefootnote}{\arabic{footnote}}
\begin{abstract}
Split Learning (SL) is a distributed deep learning approach enabling multiple clients and a server to collaboratively train and infer on a shared deep neural network (DNN) without requiring clients to share their private local data. The DNN is partitioned in SL, with most layers residing on the server and a few initial layers and inputs on the client side. This configuration allows resource-constrained clients to participate in training and inference. However, the distributed architecture exposes SL to backdoor attacks, where malicious clients can manipulate local datasets to alter the DNN's behavior. Existing defenses from other distributed frameworks like Federated Learning are not applicable, and there is a lack of effective backdoor defenses specifically designed for SL.

We present \ourname, the first defense against client-side backdoor attacks in Split Learning (SL). \ourname enables the server to detect and filter out malicious client behavior by employing circular backward analysis after a client's training is completed, iteratively reverting to a trained checkpoint where the model under examination is found to be benign. It uses a two-fold analysis to identify client-induced changes and detect poisoned models. First, a static analysis in the frequency domain measures the differences in the layer's parameters at the server. Second, a dynamic analysis introduces a novel rotational distance metric that assesses the orientation shifts of the server's layer parameters during training. Our comprehensive evaluation across various data distributions, client counts, and attack scenarios demonstrates the high efficacy of this dual analysis in mitigating backdoor attacks while preserving model utility.

\end{abstract}


%
\IEEEpeerreviewmaketitle

\section{Introduction}
\label{sec:intro}

\noindent Recently, deep neural networks (DNNs) have made significant advances, leading to the development of new training frameworks such as Large Language Models (LLMs)\footnote{Although transformers and diffusion models differ significantly from traditional DNNs, they still consist of layers and trainable parameters, making them suitable for distributed learning.}, AI-based image generation, and image recognition for self-driving cars. Concurrently, the complexity of deployed DNNs has grown rapidly to manage the ever-increasing tasks, demanding more robust computational resources. This increasing complexity presents a substantial challenge for deploying such advanced DNNs on resource-constrained devices without compromising the privacy of potentially sensitive input data.

\noindent \textbf{Split Learning (SL)} is a class of distributed learning that promises to reduce the computational load on the client side without requiring the clients to share their data. In this paradigm, the DNN's architecture is split between client and server, with the computationally intensive layers outsourced to a server~\cite{vepakomma2018split,gupta2018distributed}. Thus, it is a resource-friendly collaboration between the clients and the server, unlike federated learning~\cite{li2020federated, gao2023pcat, kairouz2021advances}. The strengths of Split Learning have been demonstrated on real-world data in medical contexts~\cite{yang2022robust}. Further potential applications include financial services and AI-based consumer services such as image editing on mobile devices, analogously to Federated Learning that is widely deployed as part of GBoard~\cite{mcmahan2017googleGboard}. In the past, different configurations for SL were developed, with the two most prevalent being the vanilla and U-shaped configurations~\cite{gao2023pcat, tajalli2023feasibility, kariyappa2023exploit, bai2023villain, zhou2023label}. In the vanilla configuration~\cite{tajalli2023feasibility,zhu2023passive}, the DNN is split into two segments at a specific "cut layer." The smaller portion of the network resides on the client side, while the larger part is on the server side. In the U-shaped configuration~\cite{zhu2023passive,gao2023pcat}, the DNN is split into three segments. The head segment, up to the first "cut layer", and the tail segment, starting from the second "cut layer", reside on the client side. The middle segment called the backbone, is outsourced to the server because it is usually composed of most of the layers and, therefore, is computationally more intensive to train. 

Training in SL proceeds sequentially, with clients queuing for sessions. In the vanilla setup, each client trains its part of the network up to the cut layer, transmitting outputs to the server, which completes the training. In U-shaped systems, the client model evaluates data up to the first cut layer during forward propagation before sending the feature vectors to the server. The server processes these vectors and returns the output for the client to complete using the tail. During backpropagation, the client calculates the loss and sends gradients to the server, which computes the backbone's gradients and sends cut layer gradients back to the client. This setup allows resource-constrained devices to train large models securely and privately~\cite{vepakomma2018split, singh2019detailed}, complying with data privacy standards like GDPR~\cite{GDPR2018} and HIPAA~\cite{HIPAA1996}. However, this setup also increases the attack surface and results in a stronger threat model, since the adversary can arbitrarily manipulate the loss. Thus, the rest of the paper focuses on the U-shape configuration. 

\noindent\textbf{Attacks on SL.} Recent works demonstrated the vulnerability of SL to multiple attacks, including data reconstruction attacks~\cite{mao2023secure}, label inference attacks~\cite{pasquini2021unleashing, fu2022label}, and backdoor attacks~\cite{bai2023villain, he2023backdoor, yu2023backdoor}. An adversary can launch a backdoor attack on the server-side~\cite{erdogan2022splitguard, yu2024chronic} or the client-side~\cite{yu2023backdoor, he2023backdoor}. 

This paper focuses on client-side backdoor attacks, as clients are more susceptible to attacks than well-protected servers. Additionally, malicious clients in SL have an advantage due to their access to data and labels, making it crucial for the server to defend against such attacks to protect benign clients. However, an efficient defense in SL is challenging because each client uses the trained model of its predecessor as a base, resulting in different starting models for each client. If a previous client was malicious, subsequent benign clients may unknowingly train on a poisoned model, {compromising also their results.}

Although in the past various defenses were proposed to mitigate attacks in other distributed learning paradigms, such as Federated Learning~\cite{fereidooni2024freqfed,fung2020FoolsGold,blanchard17Krum,bagdasaryan,rieger2024crowdguard,castillo2023fledge}, they are not applicable in SL due to the aforementioned sequential training structure.

\noindent\textbf{Our goal and contributions.}
To address the challenge of client-side backdoor attacks in SL, we introduce \ourname, a versatile and, to the best of our knowledge, the first backdoor defense for the U-shaped SL paradigm deployed at the server. \ourname operates on a rollback mechanism to employ circular backward analysis after a client's training ends, reverting to a trained state where the client under examination emerges as benign. That is, if the trained backbone at the server {shows} backdoor characteristics, this rollback mechanism reverts to a previous client's trained state and continues the examination to ensure a benign trained state is selected, bypassing all the malicious client's trained updates. 

To identify poisoned training contributions and detect malicious behavior, \ourname employs a two-fold strategy to perform the static and dynamic analysis of the server's backbone during the rollback mechanism. These metrics are based on the rationale that benign behavior contradicts the mispredictions caused by backdoor behavior. Therefore, significant changes need to be applied by the attacker to the model to introduce new behavior into the model. First, the motivation to conduct the static analysis of the server's backbone is to measure static characteristics of the backbone's parameters, such as the changes in the frequency domain, without considering changes over time during training. {An analogy for this strategy would be analyzing a music recording to see how often certain notes (frequencies)} appear without considering how the music changes over time. Here, we are only interested in the presence of the frequency of notes, not in the sequence or evolution of the music. Second, the motivation to conduct the dynamic analysis is to assess how the orientation or configuration of the backbone's parameters changes throughout training. We introduce a novel dynamic rotational distance metric that measures the extent of dynamic shifts in the backbone's values or configurations, providing insights into how the values evolve during training. {Thus, the rotational distance metric analyzes dynamic aspects, such as the flow and transition of the backbone values. Using the music analogy, the metric measures the transitions between the notes rather than their bare presence.} Therefore, \ourname uses both perspectives to analyze the clients’ models and detect backdoors. Using this analysis framework, we create a first novel robust defense against the backdoor attacks in SL. 

Two important things to note in the design of \ourname: Firstly, we do not permanently remove malicious clients from consideration. Instead, we skip the models of these clients, allowing them to be reconsidered in subsequent training epochs to ensure that misclassified benign clients are not unjustly removed. Secondly, we deploy \ourname before a client starts its training to select a benign starting point and ensure that poisoned training contributions are effectively mitigated and not used to train benign models.
In summary, our contributions include:
\begin{itemize}
\item We propose \ourname, the first defense framework designed to mitigate backdoor attacks in Split Learning (SL). \ourname accurately detects backdoor attacks and reduces their impact while minimizing harmful effects on the models' utility. We conduct static and dynamic analyses to inspect nuances in the applied model updates (\sect\ref{sec:approach-design}).

\item Our approach addresses the challenges introduced by SL's sequential structure through a circular defense mechanism applied after each client's training process. It enables the early detection of malicious client behavior and mitigates its impact before it influences the training of other benign clients (\sect\ref{sec:approach-circle}).

\item We introduce a novel rotational distance metric that measures, based on that angular displacement, how the orientation or configuration of the backbone's value at the server changes throughout training (\sect\ref{sec:approach-rotational}). This metric also captures dynamic nuances such as the orientation or rotation of the changes that provide additional information about learning objectives during the training (\sect\ref{sec:approach-rotational}).
\item We perform a deep analysis of the static changes introduced through the local training by analyzing the model updates in the frequency domain (\sect\ref{sec:approach-frequency}).
\item To evaluate the effectiveness of our defense, we developed various backdoor attacks, including poisoning and semantic backdoor attacks. These attacks were applied to datasets such as \cifar, \fmnist, \mnist,\cifarHundred, and \gtsrb across different numbers of clients, attack settings, and data scenarios. Our extensive evaluation demonstrates the effectiveness of \ourname even against defense-adapted attacks (\sect\ref{sec:eval}).
\end{itemize}

\noindent With \ourname, we make the first step towards solving an open challenge of mitigating client-side backdoor attacks on SL, significantly reducing the backdoor impact while maintaining the accuracy of the resulting model. We hope that future research continues building on top of our scheme.

\section{Background}
\label{sec:background}

\subsection{Split Learning}
\label{sec:background-sl}
\begin{figure}[t]
	\centering{
		\subfloat[Connected DNN]{
			\includegraphics[width=0.45\columnwidth]{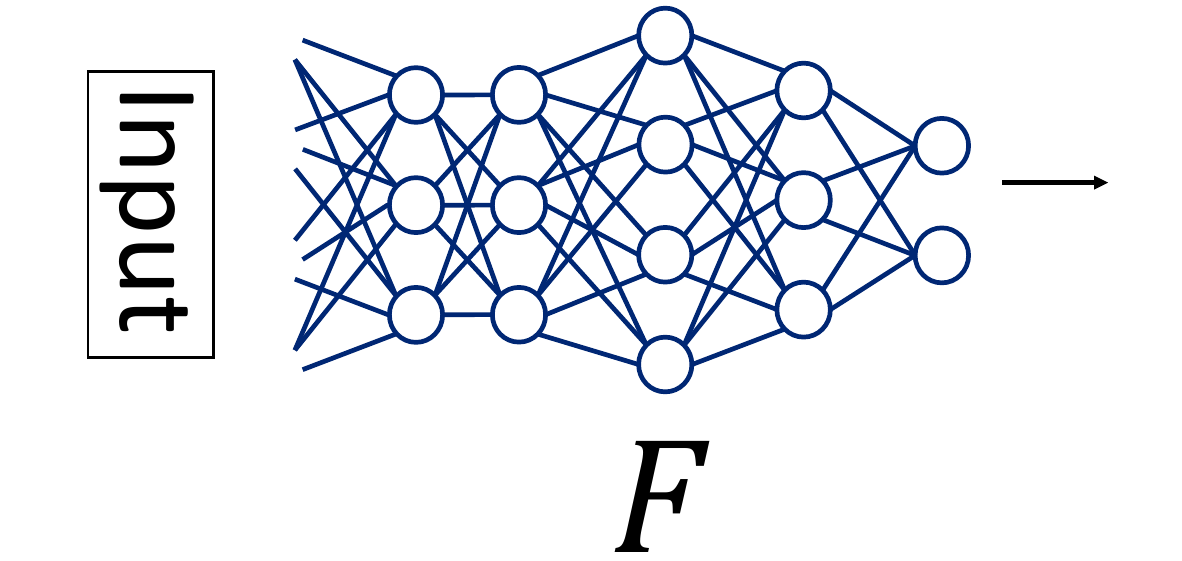}
			\label{fig:slSplit:complete}
		}
		\subfloat[DNN split into head, backbone and tail.]{
			\includegraphics[width=0.45\columnwidth]{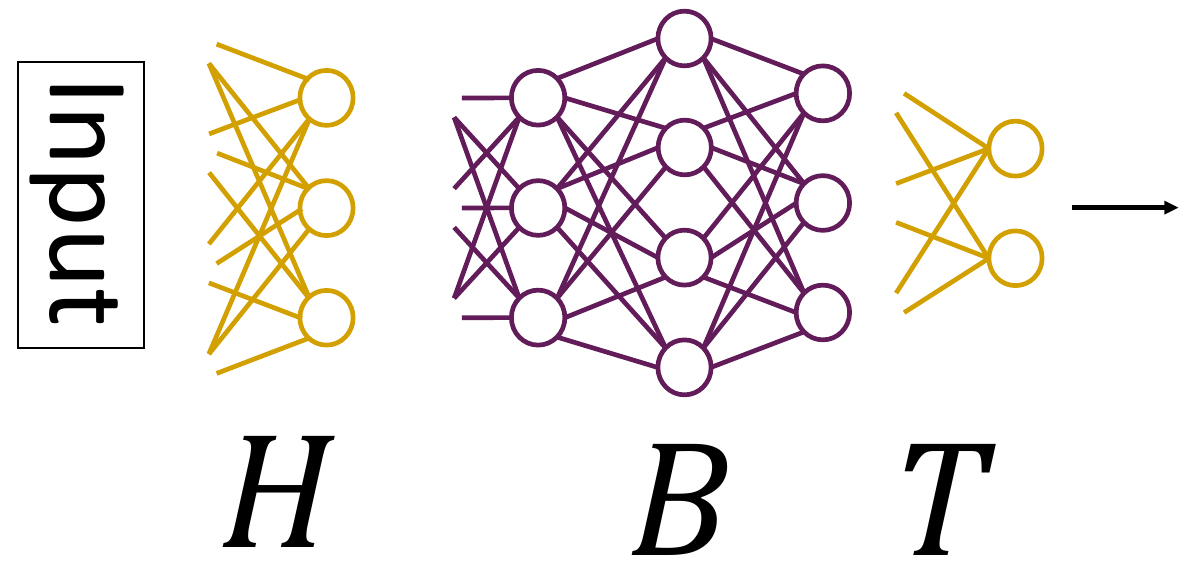}
			\label{fig:slSplit:split}
		}
	}
	\caption{Comparison of splitting the Deep Neural Network (DNN) F into head ($H$), backbone ($B$), and tail ($T$), such that $F\equiv H\circ B \circ T$. The head and tail are located on the client side, and the backbone is on the server side.}
	\label{fig:slSplit}
\end{figure}
\begin{figure*}[tb]
    \centering
    \includegraphics[width=0.9\textwidth,trim={0.cm 0.3cm 0 0.cm},clip]{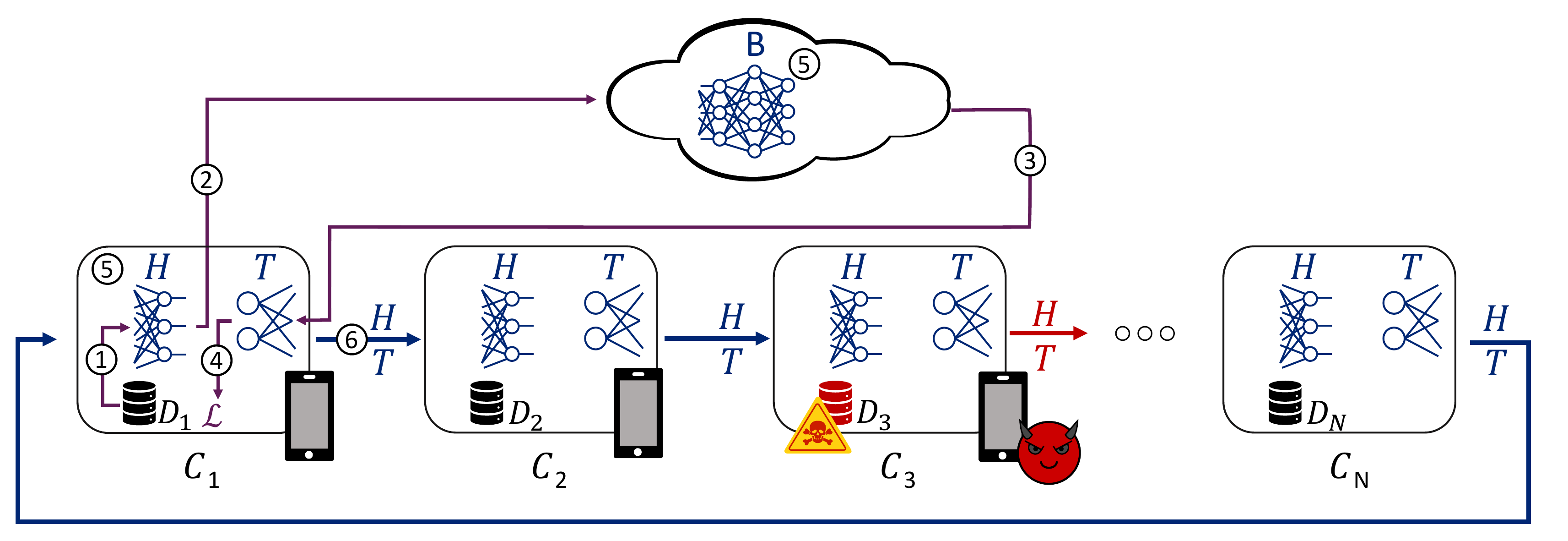}\vspace{0.5em}
    \caption{Overview of a Split Learning (SL) system that utilizes data from mobile devices but executes the computation-heavy backbone ($B$) on a cloud server, while all clients $C_1, \ldots C_{\nClients}$ provide the data $D_i$, hosts the head $T$ and tail $T$, as well as calculates the loss $\mathcal{L}$.}
    \label{fig:splitlearning}
\end{figure*}

In the U-shaped Split Learning paradigm, \nClients different participants $C_1, \dots, C_N$ jointly train a DNN coordinated by a central server. As shown in Figure~\ref{fig:slSplit}, the typical architecture of a connected DNN (Fig.~\ref{fig:slSplit:complete}) is split into three sub-models, head ($H$), backbone ($B$) and tail ($T$) ( Fig.~\ref{fig:slSplit:split}). Next, we detail some of the definitions utilized in the framework of SL. 

\noindent\textbf{Head ($H$)} resides on the client side and receives input data $D$ directly from the client's local dataset. The output of the head, the smashed data, is sent to the server during forward propagation. Later, the client receives the gradients from the server to complete the backpropagation and update the head.\\
\noindent\textbf{Backbone ($B$)} forms the central part of the model and resides on the server side. It accepts the smashed data from the head during forward propagation, further processes it, and sends it to the $T$. During backward propagation, it accepts gradients from $T$, continues the processes, and sends its output to the $H$ to complete the training.\\
\noindent\textbf{Tail ($T$)} is the last part of the model, residing on the client side. It receives intermediate features from the backbone and refines them to generate the final output predictions and backpropagates gradients to the server.

The U-shaped paradigm makes use of the function composition property that allows obtaining the exact behavior of $F(X)$ through the implicit concatenation of $H$, $B$, and $T$, as shown below:

\begin{equation}
(T \circ B \circ H)(X) \equiv F (X) \nonumber
\end{equation}

Given that the composition of functions is always associative~\cite{velleman2019prove}, we can show that feeding the head output as backbone input and giving the backbone output as tail input still represents the same behavior as the original DNN that the clients want to train.
\begin{equation}
T((B \circ H)(X)) = (T \circ B \circ H)(X) \nonumber
\end{equation}

In a standard training scenario, as depicted in Figure~\ref{fig:splitlearning}, client $i$ passes its data through its head (step 1), then gives the output of the head to the server (step 2), passes it through its backbone, and sends the output back to client $i$ (step 3), who feeds this last output to its tail and calculates the loss using the ground-truth labels (step 4). During backpropagation (step 5), the client first calculates the gradients for the tail and passes the computed gradients to the server, which again computes the {gradients} for the backbone using the information provided by client $i$. The server then applies backpropagation and passes the gradients to client $i$, who applies backpropagation {to obtain the gradients for} the head. After these steps, client $i$ shares the resulting tail and head with client $i+1$ (step 6), who uses them as starting weights for its training step.
As the server does not have direct access to the client's head and tail nor to the client's data, SL enables resource-constrained devices to train and apply large models in a secure and privacy-preserving~\cite{vepakomma2018split, singh2019detailed,gupta2018distributed}.

\section{Problem Setting}
\label{sec:problem}

\noindent In this section, we describe the considered system (\hyperref[sec:problem-system]{\sect\ref{sec:problem-system}}) and characterize the threat model (\hyperref[sec:problem-advmodel]{\sect\ref{sec:problem-advmodel}}), before describing inherent challenges in SL in mitigating backdoor attacks (\hyperref[sec:problem-challenges]{\sect\ref{sec:problem-challenges}}). In the appendix, we provide a high-level overview of DNNs (App.~\ref{app:dnn}), poisoning attacks (App.~\ref{app:background-poisoning}), and the eligibility of frequency transformations to detect backdoors (App.~\ref{app:background-dct}).

\subsection{System Setting}
\label{sec:problem-system}
In the rest of this paper, we consider a system consisting of \nClients clients holding private datasets that are not shared with other parties. Coordinated by a central server \server, they use the SL framework to train a DNN on their private datasets collaboratively. An example of such a system is visualized in Fig.~\ref{fig:splitlearning}. Following existing literature~\cite{gao2023pcat,liu2024similarity,poirot2019split,zhu2023passive}, we focus on a U-shaped SL configuration that splits the DNN into three parts (Head - $H$, Backbone - $B$, Tail - $T$), where $H$ and $T$ are located on the client side, whereas $B$ is executed on the server, as {described in \sect\ref{sec:background-sl}.} Since the last part of the DNN ($T$) is on the client side, the clients are also responsible for the loss calculation. Once the training is finished, the client signs the current model and forwards it with the previous clients' models to the next client for the following training iterations.

For example, the clients could be low-performance devices such as smartphones that want to train and perform inference on a large DNN (e.g., a Large Language Model) without revealing their private training or inference data. The U-shape enables them to outsource the computation-intensive part to a high-performance server without violating the data's privacy.

\subsection{Adversary Model}
\label{sec:problem-advmodel}
\noindent 
We consider an adversary \adversary that aims to inject a backdoor into the collaboratively trained model $F$. Thus, the backdoored model $F^*$ shall predict a specific, adversary-chosen target label $L_{\adversary}$ when the model receives input $x^*$ containing the trigger $R$. When given a clean input without the trigger, the backdoored model must generate the correct prediction to prevent the backdoor from being detected. More formally:
\begin{equation}
F^*(x) = \begin{cases}
L_{\adversary} &R\in x \\
F(x) & R\not\in x
\end{cases}
\end{equation}

Notably, for a triggered input sample $x^*$, the clean prediction differs from the backdoor target label $F(x^*) \not= L_{\adversary}$. 

\adversary is assumed to have complete control over {one or several} malicious clients and can, therefore, arbitrarily manipulate the input data and labels. Further, due to the considered U-shape architecture, \adversary can arbitrarily change the local head and tails, the smashed data and gradients sent by the server, and the loss function. Aligned with existing work on mitigating backdoor attacks in distributed learning~\cite{shen16Auror,blanchard17Krum,krauss2023mesas,kumari2022baybfed,rieger2024crowdguard}, we assume \adversary to control at most $\nicefrac{\nClients}{2}-1$ clients. Additionally, we assume \adversary knows the defense mechanism deployed on the server side. Thus, \adversary can constrain the training loss utilizing the metrics used by the defense mechanism.

In the following, we will focus on the attacks that malicious clients perform. The server has an intrinsic motivation to produce a well-trained and effective model, as its reputation is based on the quality of the resulting model. Additionally, while the client devices are mostly anonymous, as they are just mobile devices, the server is identifiable and accountable for its behavior. Since existing literature already investigated the problem of defending against backdoor attacks that are conducted by malicious servers~\cite{erdogan2022splitguard}, we will consider these attacks to be out of the scope of this paper.

\subsection{Objectives and Challenges}
\label{sec:problem-challenges}
\noindent 

This section details the objectives of a defense that aspires to prevent backdoor attacks and the challenges encountered in designing such a defense in the framework of SL. 

An effective and practical backdoor defense aims to fulfill the following security objective: \\

\noindent\textbf{O1 Prevent Backdoor Attacks:} The primary requirement of an effective defense strategy is to efficiently prevent malicious clients from injecting a backdoor (as {described} in App.~\ref{app:background-poisoning}) into the trained model.\\

However, to ensure that the defense is practical and does not render the resulting model unusable, the defense must {also} fulfill the following functional requirement: \\

\noindent\textbf{O2 Preserve Model's Utility:} The defense must not negatively affect the accuracy of the model on the benign main task (Main Task Accuracy, MA).\\

This dual focus guarantees the core functionalities, resilience, and reliability.

Compared to centralized or distributed learning settings such as Federated Learning, poisoning defenses in split learning face several unique challenges.\\
\textbf{\challenge{1}: No Data Access:} The training data are located on the client side, preventing the server from inspecting them to detect manipulated training samples. Especially for scenarios that involve sensitive data, it is impractical to assume that clients share their data with the server. Therefore, the server can detect the backdoor only by analyzing the model updates.\\
\textbf{\challenge{2}: Non-Comparable Client Models:} Another challenge for detecting poisoning attacks in SL is that the models (updates) of different clients cannot be directly compared to each other. In SL, clients train in sequential order, and each client uses the trained model of its predecessor as the base model for its training. Therefore, the resulting models of two clients will always differ, even if both clients used the same data. Additionally, strategies frequently adopted in Federated Learning~\cite{fung2020FoolsGold} to compare the models' updates cannot be transferred to SL. Due to the non-convex training process of DNNs, two clients with the same data might obtain different model updates if they start training from different base models.\\
\noindent\textbf{\challenge{3}: Sequential Training:} Sequential training also poses a significant challenge in handling detected poisoned models because they affect the training results of following clients. In other distributed learning settings, such as Federated Learning, the defender typically waits until each client has trained its local model before analyzing the model updates and excluding suspicious ones~\cite{fereidooni2024freqfed,blanchard17Krum}. However, the sequential training of SL means that if a poisoned model is detected at the end of a training round when every client has finished its training, the models of all clients that followed the malicious clients would need to be discarded. Since these clients used a poisoned model as their base, their models are likely also to contain the backdoor, making them unusable for the server.

\begin{figure*}[tb]
    \centering
    \includegraphics[width=0.55\textwidth,trim={2.cm 0.cm 3.cm 0.cm},clip,angle=270]{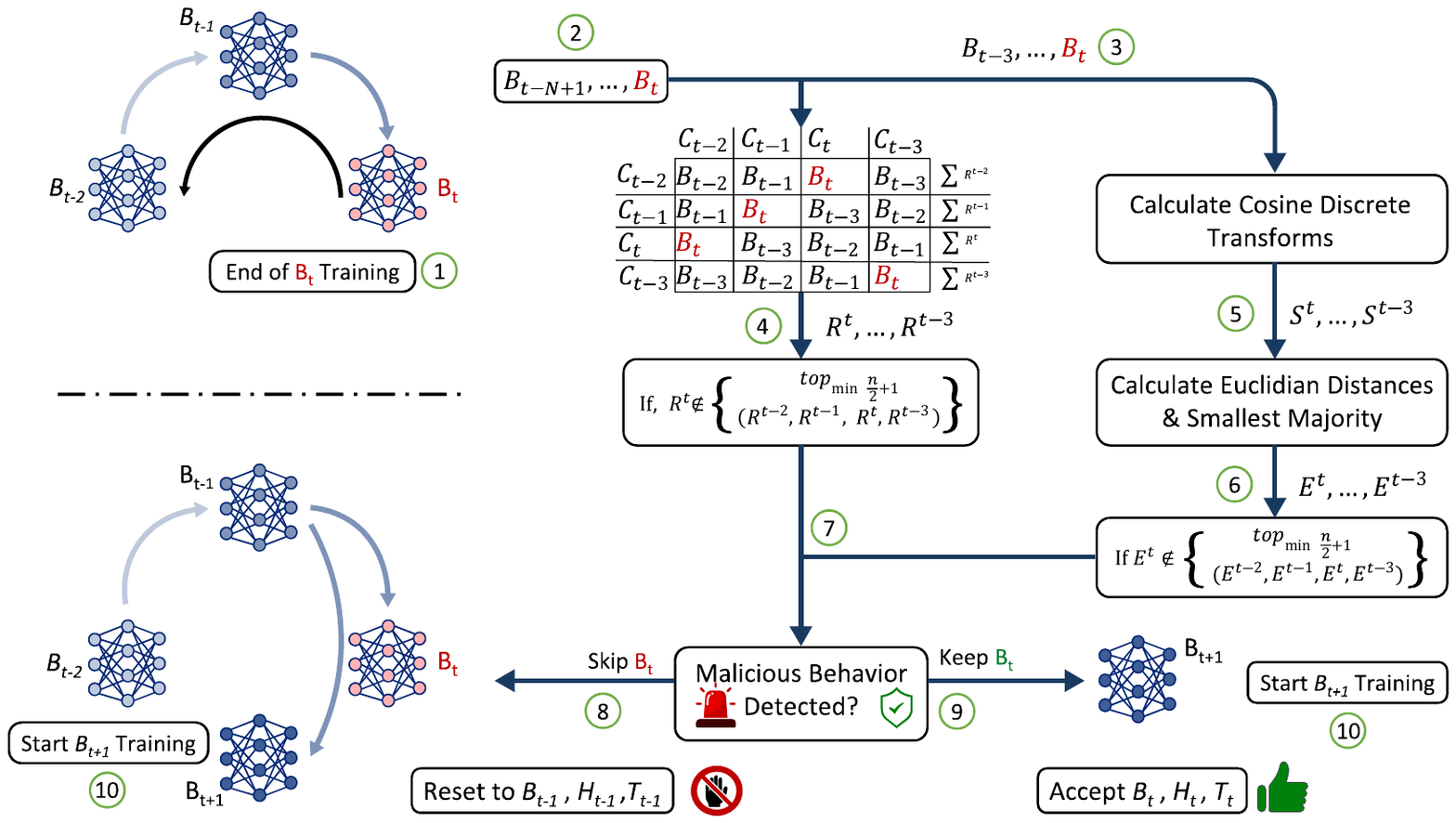}
    \caption{Workflow of \ourname to skip or poisoned models based on an analysis of the models in the frequency domain and their rotational displacement. The workflow is shown for an example scenario consisting of 4 clients.}
    \label{fig:defense-modelSkipping}
\end{figure*}

In addition, even if it would be practical to repeat the training using a different base model assumed to be benign, a sophisticated adversary might alternate between benign and malicious behavior to fool the server and avoid being removed from the pool of assumed benign clients. Then, when the training process is repeated, the adversary could try to introduce the backdoor using a client who had previously behaved inconspicuously. Therefore, even if the training is repeated, the defense must also be applied during this repetition. {On the other side, in the case of the absence of any attack, the defense must be able to accept all models if no poisoning attack is detected. Otherwise, the defense would cause an endless loop of repeating the training after seemingly malicious clients are detected and excluded malicious clients, rendering existing outlier-detection-based techniques~\cite{blanchard17Krum,shen16Auror,munoz19AFA} impractical for Split Learning.}

\section{Rationale for Static and Dynamic Analysis}
\label{sec:approach-rationale}
In the following, we elaborate on the intuition behind the two-fold analysis that is employed to efficiently detect poisoned models.

\noindent\textbf{Static Analysis: }
The purpose of the frequency analysis is to measure fine-granular static changes that are performed in the backbone during the clients' training. {Previous} research~\cite{Rahaman2019spectralbias,xu2019training} has shown that in the early stages of the training, mostly the lower components of the models' frequency representation change. Only with progressing convergence do the high frequencies start to change significantly. Therefore, the low-frequency components are especially affected when a model is trained for new behavior. However, when injecting a backdoor that was trained on benign data in advance by earlier clients, the backdoor behavior is in contradiction with the benign model's behavior as the backdoor target label differs from the benign prediction (see \sect\ref{sec:problem-advmodel}). Therefore, significant adaptions of the model's behavior are necessary, resulting in high changes for the low-frequency components.

To perform this analysis, we first transform each model update into the frequency domain using the Discrete Cosine Transform (DCT) (App.~\ref{app:background-dct}). The DCT helps break down the model's updates into their frequency components, allowing unusual changes that indicate backdoor behavior to be spotted. Specifically, we use the two-dimensional DCT (2-D DCT) because of its energy compaction property, which means it stores the most important information in the low-frequency components. This characteristic helps detect significant changes, such as those introduced by backdoor attacks. Additionally, the DCT is computationally advantageous over the Discrete Fourier Transform (DFT) because its output is always in real numbers, making it simpler and faster to compute. By transforming the model updates with the DCT and then calculating the pairwise Euclidean distances between these frequency representations, we can effectively detect anomalies that suggest the presence of backdoor behavior.

To convert a backbone's distances to all other backbones into a score, we sum up the distances to the closest $\nicefrac{n}{2}+1$ other models. As described in \sect\ref{sec:problem-advmodel}, we assume a majority of clients to be benign. Thus, a model that does not belong to the majority is considered malicious during the current rollback.

Thus, by considering only the $\nicefrac{n}{2}+1$ smallest distance values, we prevent \adversary from manipulating the score calculation, e.g., by providing manipulated models (so-called canaries) that increase the scores of benign models to make the selection of a regular poisoned more likely. By considering only the smallest majority of scores is considered, we ensure that \adversary cannot increase the score for benign models.

\noindent\textbf{Rotational Analysis: }
During training, backdoor attacks can cause significant and unusual changes in the orientation or configuration of the model's parameters. To detect such anomalies, \ourname employs a dynamic analysis by computing the rotational distance between pre-trained (historical) server backbone updates. This rotational metric captures the extent of changes in the model's parameter space, providing a robust measure of how the backbone evolves.

The rotational distance metric is designed to quantify the orientation shifts in the model's high-dimensional parameter space. Unlike static analysis, which focuses on the magnitude and plain difference for parameter updates, dynamic analysis focuses on the direction and trajectory of these updates. By analyzing the rotational distance between different backbones, \ourname can identify deviations from typical training patterns indicative of backdoor behavior. This can be seen as an interpolation of the path that the optimizer took during training from the base model $B_{t-1}$ to the trained model $B_{t}$, thus revealing information about the training objective.

\ourname computes the rotational distance metric $R_{D}$ (more details in Section \ref{sec:approach-rotational}) for each server's backbone. The reasons for utilizing it in the backdoor detection are three-fold. First, $R_{D}$ analyzes directional changes. The rotational distance metric is sensitive to the direction of parameter updates. Backdoor attacks typically introduce abrupt and significant directional changes to implant malicious behavior, which can be detected by observing large rotational distances. Second, $R_{D}$ is robust to scaling. By focusing on the orientation rather than the magnitude, the rotational distance metric is robust to variations in the scale of parameter updates, which might occur due to changing client-controlled hyperparameters such as the learning rate, optimization algorithm or loss function. Third, $R_{D}$ complements the static analysis. While static analysis detects anomalies based on the magnitude and difference of updates in the frequency domain, dynamic analysis adds a layer of scrutiny by examining the trajectory of these updates. Together, they cover both perspectives and provide a comprehensive defense mechanism.

Another important aspect of \ourname is the circle-wise analysis that ensures using the latest benign-detected model and, therefore, skipping the malicious clients. This prevents benign clients from training on poisoned models. This approach strengthens the security of the training process, ensuring model reliability as we do not remove the benign clients that have been misclassified as malicious.

\section{\ourname}
\label{sec:approach}
\begin{figure}[tb]
    \centering
    \includegraphics[width=0.95\columnwidth,trim={7.12cm 0.2cm 3.2cm 0.1cm},clip]{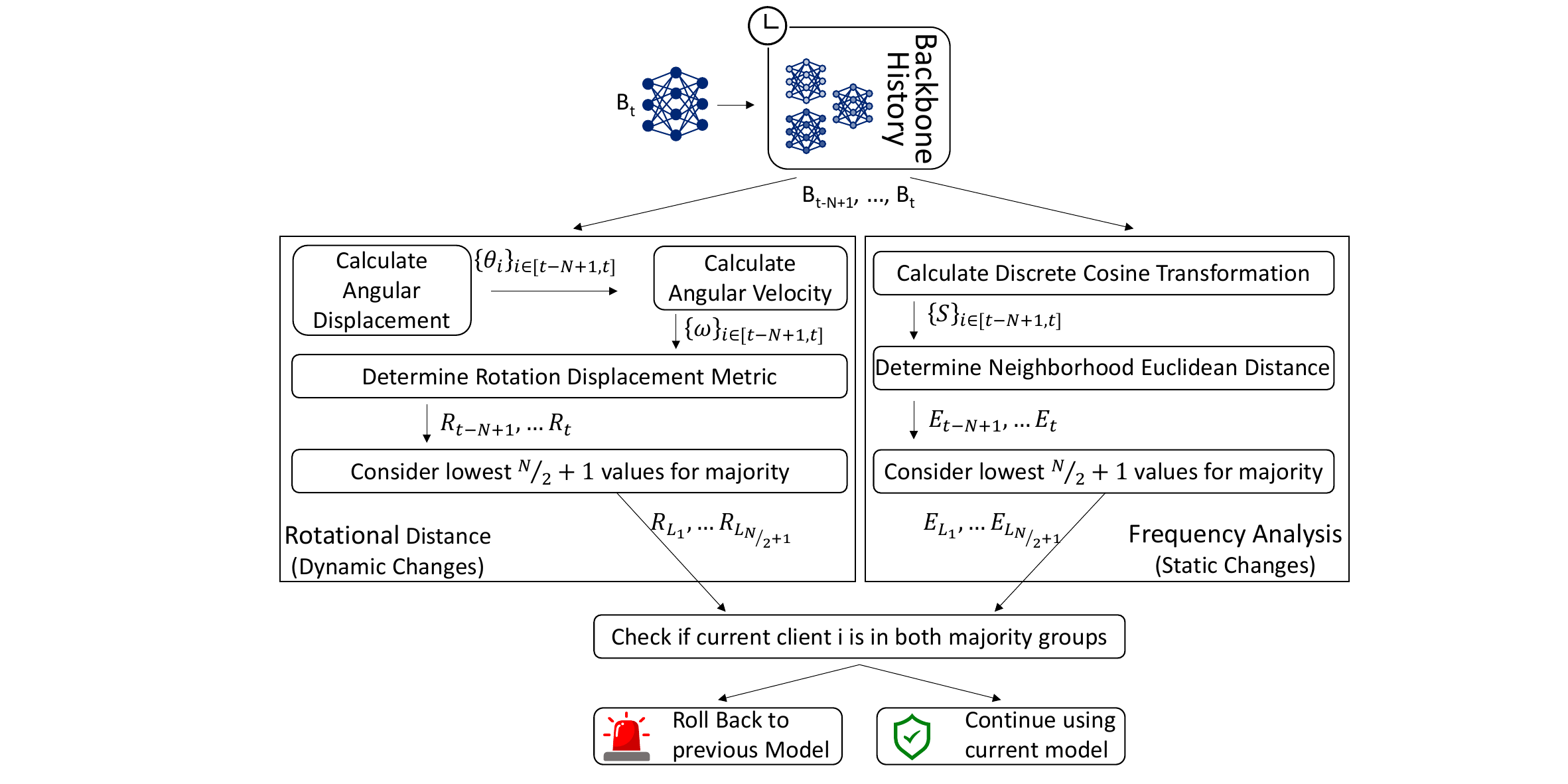}
    \caption{Overview of \ourname, using the latest backbone model $B_t$ and previous backbones $B_{t-\nClients+1}\ldots B_{t-1}$ to determine Rotation Displacement Metric values $R_{t-N+1}, \ldots, R_t$ and the Euclidean Distance Neighborhood Scores, before returning the index of most recent backbone $L_i$ being among $\nicefrac{\nClients}{2}+1$ lowest values $R_{L_1}, \ldots, R_{L_{\nicefrac{\nClients}{2}+1}}$ and $E_{L_1}, \ldots, E_{L_{\nicefrac{\nClients}{2}+1}}$.}
    \label{fig:overview}
\end{figure}

\noindent In the following, we describe the high-level design of \ourname (\sect\ref{sec:approach-design}) and the underlying intuition (\sect\ref{sec:approach-rationale}) before elaborating on its model-skipping mechanism (\sect\ref{sec:approach-circle}): frequency analysis to detect static changes (\sect\ref{sec:approach-frequency}), and rotational distance analysis to detect dynamic changes (\sect\ref{sec:approach-rotational}). The overall workflow of \ourname is visualized in Fig.~\ref{fig:defense-modelSkipping} and formalized in Alg.~\ref{alg:ouralg}. {Fig.~\ref{fig:overview} shows the analysis process in more detail.}

\subsection{High-level Design}
\label{sec:approach-design}
\ourname is deployed on the server and analyzes the model's backbone parameters updates. Thus, \ourname is executed {before each client's training starts to select a benign starting point}. The high-level process is shown in Fig.~\ref{fig:defense-modelSkipping}. Each time the backbone parameters are {updated after a client finishes its training and before the next client starts its training} (step 1 in Fig.~\ref{fig:defense-modelSkipping}), \ourname analyzes the updated backbone for backdoor behavior (steps 2 - 9 in Fig.~\ref{fig:defense-modelSkipping}). If no malicious behavior is detected, the training process continues by training the next client using the latest model of the previous client as the base model (steps 9 and 10 in Fig.~\ref{fig:overview}). 

However, if backdoor behavior is detected, the rollback mechanism iterates through the previous clients' training outputs to identify the backbone's latest benign state. Thus, this rollback mechanism reverts to the previous server's backbone state and continues the examination to ensure a benign trained state is indeed selected, bypassing all the malicious client’s trained updates (step 8 in Fig.~\ref{fig:overview}). In this case, the current backbone is replaced by the identified latest benign checkpoint, and the next client is instructed to train using the respective head and tail associated with the benign backbone found (step 10 in Fig.~\ref{fig:overview}).

We employ a two-fold analysis strategy to detect the benign state of the server's backbone and detect poisoned updates. First, we perform a static analysis to analyze the Euclidean distances of the backbones' frequency representation (steps 5 and 6 in Fig.~\ref{fig:overview}). Second, we perform a dynamic analysis to quantify the evolution of the backbone's values using our novel rotational distance metric (step 4 in Fig.~\ref{fig:overview}). Both metrics are based on the observation that the backdoor behavior contradicts the benign behavior. While different benign clients either have similar training behavior in the case of \iid data or orthogonal behavior in the case of \nonIid data, the backdoor aims to make the model mispredict a wrong label (see \sect\ref{sec:problem-advmodel}), to the backdoor target. Thus, the malicious client needs to change parts of the benign training and train a new backdoor behavior embedded in the model, resulting in large distances for the leveraged metrics.

We use both metrics to determine the benign state of the server's backbone, as each investigates a different perspective. The distances of each metric are used to determine a score for each model that indicates the alignment of the model's training objective. A high score indicates that the current model's training objective was in contradiction with the behavior of a majority of other clients. \ourname considers a model to be benign if, for both metrics, the model's scores are small, thus within the respective sets of smallest set $\nicefrac{\nClients}{2}+1$ of existing score values (see lines 17-23 in Alg.~\ref{alg:ouralg} and step 7 in Fig.~\ref{fig:overview}). Since the majority of models are assumed to be benign (see \sect\ref{sec:problem-advmodel}), a model that does not belong to the majority is considered malicious during the current rollback.

\subsection{Circular Benign Model Identification}
\label{sec:approach-circle}

After each client $i$ completes training in their respective communication round $t$ with the server (lines 8 - 9 in Alg.~\ref{alg:ouralg}, step 1 in Fig.~\ref{fig:overview}), we store the server's backbone $B_t$ in a FIFO array (step 2 in Fig.~\ref{fig:overview}. This process is repeated for each training step $t$, ensuring we retain the last \nClients models. Once client $i$ finishes training, we compute the deployed metrics (Rotational Distance and Euclidean distance of frequency representation) for all backbone models $B_{t-\nClients+1}, \ldots, B_t$ stored in the FIFO array (lines 10-16 in Alg.~\ref{alg:ouralg} and steps 3-6 in Fig.~\ref{fig:overview}).

Next, we determine the benign checkpoint $B$ (lines 17-25 in Alg.~\ref{alg:ouralg}). This process determines whether the backbone being examined shows backdoor behavior. For this, the server checks for each metric if the model's score is within the smallest $\nicefrac{\nClients}{2}+1$ of existing score values (lines 17-19 in Alg.~\ref{alg:ouralg} and step 7 {in} Fig.~\ref{fig:overview}, see also \sect\ref{sec:approach-rationale}). Alternatively, this process determines how many clients to skip in backward direction to reach a benign backbone checkpoint, i.e., obtain $B$ and the index of the corresponding optimal head ($H$) and tail ($T$) of the client determined showing benign behavior (step 8 in Fig.~\ref{fig:overview}). Afterward, the next client is informed which head and tail to use as base model for it's training (step 10 in Fig.~\ref{fig:overview}).

It should be noted that we describe the circular benign model identification for a system where only the backbone is held on the server, while the head and tail always reside on the client side and are forwarded from each client to its successor. As described in the system setting (\sect\ref{sec:problem-system}), each client signs its training result to allow the succeeding clients to verify the origin of the respective used head and tail. This ensures that the server never gets access to the head and tail, leading to improved privacy. However, \ourname can be straightforwardly adapted to scenarios where each client sends the {head} and tail after the training to the server, and the server forwards them to the next clients.

\subsection{Frequency Variation Computation}
\label{sec:approach-frequency}

After each client $i$ finishes training, we use the Discrete Cosine Transformation (DCT) to determine the frequency representation of each model and select the low frequencies (line 10 in Alg.~\ref{alg:ouralg} and step 5 in Fig.~\ref{fig:overview}). Then, we compute the Euclidean distance in the frequency domain for the server's backbone parameters (line 11 in Alg.~\ref{alg:ouralg}). Let $\mathbf{B}_t$ represent the backbone parameters of the server at the training step $t$. The low-frequency component of the backbone's update $S_t$ is computed as:
\begin{equation}
    S_t = \text{DCT}_{\text{low}}(\mathbf{B}_{t} - \mathbf{B}_{t-1})
\end{equation}

where $\text{DCT}_{\text{low}}(\mathbf{B}_t)$ denotes only the low-frequency components of the Discrete Cosine Transform (DCT) applied to the backbone's update $\mathbf{B}_{t} - \mathbf{B}_{t-1}$. We then calculate for each frequency representation $S_t$ the Euclidean distances to each other frequency representation (step 6 in Fig.~\ref{fig:overview}):

\begin{equation}
    ed_{t,i} = \| S_t - S_i \|_2
\end{equation}

The frequency score $E_t$ of the model $B_t$ is then calculated as the sum of the distances to the frequency representation of the $\nicefrac{\nClients}{2}+1$ closest other models (line 12 in Alg.~\ref{alg:ouralg} and step 6 in Fig.~\ref{fig:overview}).

The DCT converts the backbone parameters $\mathbf{B}_t$ into their frequency domain representation. This transformation allows us to analyze how the frequency components of the backbone parameters change over time. The Euclidean distance $D_t$ quantifies the pairwise differences between the frequency representations of the backbone parameters for each client $i$ and $j$ in ($\mathbf{B}_{t-\nClients+1}, \mathbf{B}_{t}$). Larger distances may indicate significant changes in the backbone's frequency characteristics, which could be indicative of injecting contradicting behavior, being typical for backdoor attacks. This static analysis provides insights into the stability and consistency of the server's backbone parameters over multiple training steps, helping to detect and mitigate potential security risks in the SL framework.

\subsection{Measuring Rotation Distances}
\label{sec:approach-rotational}
In this section, we describe the computation of the rotational distance metric (step in Fig.~\ref{fig:overview}) that captures the dynamic changes in the server's backbone's update for each client's training. As mentioned in \sect\ref{sec:approach-circle}, to determine the checkpoint of the benign behavior, we keep track of the \nClients historical backbone updates and compute the pairwise differences between \RDM of the backbone parameters for each client $i,j \in \{\mathbf{B}_{t-\nClients+1}, \mathbf{B}_{t}\}$ after training the current client $i$. \RDM computation consists of three steps: First, we compute the angular displacement of the parameter values in the backbone (line 13 in Alg.~\ref{alg:ouralg}). Second, we compute the angular velocity to quantify the rate of change in the angular displacement. Lastly, we compute the rotational frequency to identify the frequency of these directional changes.

\noindent \textbf{Angular Displacement Computation. }The angular displacement $\theta(t)$ measures the orientation change between successive backbone updates. To inject {contradicting backdoor behavior, the attack introduces significant directional changes,} causing $\theta(t)$ to deviate from normal training patterns. Thus, given the current client $i$ and its backbone model $B_{i}$, we compute the angular displacement that measures a rotating object's change in orientation (line 13 in Alg.~\ref{alg:ouralg}).

Let $\mathbf{B}_t$ be a backbone at step t. The angular displacement $\theta(t)$ is given by\footnote{In an earlier version of this manuscript, by accident, "arccos" instead of "arctan" was written.}:
\begin{equation}
    \theta(t) = \arctan \left(B_t \right)
\end{equation}

We compute $\theta(t)$ for each client $i$  in ($\mathbf{B}_{t-\nClients+1}, \mathbf{B}_{t}$) after training the current client (see App.~\ref{app:rotational} for details). This allows to obtain the angular position at each time point. The angular displacement $\theta(t)$ measures the change in orientation between successive backbone updates. Backdoor attacks often introduce significant directional changes, causing $\theta_t$ to deviate from normal training patterns.

\noindent \\
\textbf{Angular Velocity Computation. }The rate of change of the angular displacement over time, or angular velocity $\omega(t)$, is given by:
\begin{equation}
   \omega(t) = \frac{\theta(t) - \theta(t-1)}{\Delta t}
\end{equation}

where $\Delta t$ is the time interval between successive updates. The angular velocity $\omega(t)$ quantifies the rate of change in $\theta(t)$ (see line 14 in Alg.~\ref{alg:ouralg}). High angular velocities $\omega(t)$ indicate rapid shifts in the parameter space, which are uncommon during regular training but typical of backdoor insertion attempts.\\

\textbf{Rotational Distance Metric (\RDM) Computation. } {At the end, we compute the} rotational displacement metric $R_{D}$, which is also the rotational frequency that captures the frequency of these directional changes (see line 15 in Alg.~\ref{alg:ouralg}). \RDM is defined as:

\begin{equation}
   R_{D} = \frac{\omega(t)}{2\pi} \label{eq:rd}
\end{equation}

We divide $\omega(t)$ by $2\cdot \pi$ because one complete revolution (or cycle) around a circle corresponds to an angle of $2\cdot \pi$ radians. So, dividing by $2\cdot \pi$ allows us to convert $\omega$, measured in radians per second, into the rotational frequency, typically measured in cycles per second or Hertz (Hz).

The rotational frequency $R_{D}$, derived from $\omega(t)$, identifies the frequency of these directional changes. A high $R_{D}$ suggests frequent and abrupt alterations in parameter orientation, which is characteristic of backdoor attacks that aim to alter model behavior.\\

The rotational distance metric allows capturing dynamic changes in the orientation or configuration of the backbone during the local training of clients. It primarily measures angular changes between gradient vectors (geometrically). As we show in the following sections, this makes the metric highly effective for identifying deviations of poisoned models. A major strength of the metric is its ability to analyze the rotational changes over time, making it resilient to small, random perturbations or noise introduced in the gradients. Hence, it is able to discern the changes between the malicious updates even if the noise is systematic and mimics the patterns introduced by a poisoning attack. This makes the metric robust against manipulations to hide the poisoned gradients.

After the DCT and \RDM computation, we use them to compute the benign checkpoint to superimpose the checkpoint's backbone model $B^{*}$ to the current backbone state $B_{i}$ at the server and transfer the corresponding optimal head ($H^{*}$) and tail ($T^{*}$) of the client at the benign checkpoint to the next client for its training (to use as the base model).

\begin{algorithm}[tb]
\caption{\ourname}
	\scalebox{0.75}{
	\begin{minipage}{1.25\columnwidth}
\small
    \label{alg:ouralg} 
    \begin{algorithmic}[1]       
        \State \textbf{Input:} $\nClients$, $R$, $B_0$, $H_0$, $T_0$, \text{clients} \Comment{ \nClients is the number of clients, $R$ the number of training rounds,  $H_0, B_0, T_0$ build the initial model, a FIFO list of \nClients clients}
        \State \textbf{Output: $B^*$, $H^*$, $T^*$} 
        \Comment{updated backbone, heads and tails}
        
        \Statex
        \Function{SmallestMajority}{$v_1, \ldots, v_N$}
        \State \Return sorted($v_1, \ldots, v_N$)[1,\ldots,$\nicefrac{N}{2}+1$]
        \EndFunction
        \Statex

        \State $H, B, T \gets H_0, B_0, T_0$ 
        \For {each training step $t \in [1, R \cdot \nClients]$}
        \Statex\Comment{Perform Training}
        \State $\text{current\_client} \gets \text{clients}[t \text{ mod } \nClients]$
        \State $H_t^{*}, B_t^{*}, T_t^{*} \gets$ \Call{Train}{$\text{current\_client}, H^{*}, B^{*}, T^{*}$}
        
        \Statex \Comment{Measure Distance in Frequency Domain}
        
        \State $\forall i \in \{t-\nClients+1, \ldots, t\} S_{i} \gets \text{DCT}(B_i - B_{i-1})$
        
        \State $\forall i,j \in \{t-\nClients+1, \ldots, t\} ed_{i,j} \gets \text{EuclideanDistance}(S_i, S_j)$
        \State $\forall i \in \{t-\nClients+1, \ldots, t\} E_i \gets \sum \text{SmallestMajority} (ed_{i,t-\nClients+1}, \ldots,ed_{i,t} )$
        
        \Statex \Comment{Calculate Rotational Distance}
        \State {$\forall i\in \{t-\nClients+1, \ldots, t\} \theta_{i} \gets \text{AngularDisplacement}(B_i)$}
        \State {$\forall i \in \{t-\nClients+1, \ldots, t\} \omega_{i} \gets \text{AngularVelocity}(\theta_{i})$}
        \State $\forall i \in \{t-\nClients+1, \ldots, t\} \text{RD}_i \gets \text{RotationalDisplacement}(\omega_i)$
        \State {$\forall i \in \{t-\nClients+1, \ldots, t\} R_i \gets \sum \text{SmallestMajority} (i, RD_{t-\nClients+1}, \ldots, RD_{t} )$}
        
        \Statex \Comment{Determine Benign Majority}
        \State $\text{frequency\_majority} = \text{arg\_sort}(E_{t-\nClients+1}, \ldots, E_t)[1,\ldots,\nicefrac{N}{2}+1]$        
        \State {$\text{rotation\_majority} = \text{arg\_sort}(R_{t-\nClients+1}, \ldots, R_t)[1,\ldots,\nicefrac{N}{2}+1]$}
        \State $\text{benign\_majority}\gets \text{rotation\_majority} \cap \text{frequency\_majority} $
        \Statex\Comment{Determine Latest Benign Model}
        \For {each previous client $c \in [t, \max(t - \nClients + 1, 1)]$} 
        \If{$c \in \text{benign\_majority}$}
            \State $H^{*}, B^{*}, T^{*} \gets H_c, B_c, T_c$ 
            \State \textbf{break}
        \EndIf
        \EndFor

        \EndFor 
        \State \Return $H^{*}, B^{*}, T^{*}$
       
    \end{algorithmic}
    \end{minipage}}
\end{algorithm}

\section{Evaluation}
\label{sec:eval}
This section presents a comprehensive evaluation of \ourname across various {types of backdoor attacks}, showcasing our defense mechanism's ability to maintain backdoor accuracy below 5\%. In App.~\ref{app:distanceScores}, we analyze the effectiveness of the combination of static and dynamic perspectives. {In App.~\ref{app:further} we evaluate further non-backdoor attacks.}

\subsection{Experimental Setup}
\label{sec:eval-setup}
\noindent 
\noindent\textbf{Metrics}: Our evaluation of \ourname leverages the following key metrics.

\noindent\textit{Backdoor Accuracy (BA)} indicates the model's accuracy concerning malicious tasks. It quantifies the fraction of the trigger set for which the model generates predictions aligned with the attacker's objectives. The attacker aims to maximize BA, while \ourname strives to minimize it.\\
\textit{Main Task Accuracy (MA)} measures the model's accuracy for benign tasks. It reflects the percentage of clean inputs for which the model delivers accurate predictions. The adversary aims to reduce the impact on MA to diminish the likelihood of detection. An essential requirement of \ourname is not to significantly affect the MA.\\

\noindent\textbf{Datasets.}
Aligned with existing work on SL~\cite{yu2024chronic,he2023backdoor,pasquini2021unleashing,yu2023backdoor,bai2023villain,tajalli2023feasibility,gao2020end,erdogan2022splitguard,thapa2021advancements,fu2023focusing}, we leveraged five datasets (\cifar, \fmnist and \mnist, \gtsrb, \cifarHundred)  to perform our experiments:\\

\noindent\textit{\cifar} consists of \numprint{50000} training and \numprint{10000} test images of size $32\times32$ pixels, showing objects and animals belonging to 10 different classes~\cite{krizhevsky2009learning}. As DNN, we use the widely adopted ResNet-18 architecture.\\
\textit{\mnist} consists of \numprint{60000} training and \numprint{10000} test grayscale images showing handwritten digits. Aligned with recent work on distributed learning, we implemented a Convolutional Neural Network (CNN), as described by Cao \etal~\cite{cao2021provably}.\\
\textit{\fmnist} is composed of \numprint{60000} training and \numprint{10000} test images, each sized 28x28 pixels, depicting various types of clothing across 10 classes~\cite{fmnist}. As DNN we use also the CNN described by Cao \etal~\cite{cao2021provably}.\\
\textit{\cifarHundred} consists similar to \cifar of \numprint{50000} training and \numprint{10000} test images but is categorized into 100 classes. Due to the high number of labels, which are significantly larger than the considered number of clients. As DNN, we used Wide-ResNet50, being pretrained for ImageNet dataset and replaced the final layer, while the training included all layers.\\
\textit{\gtsrb} consists of \numprint{39000} training and \numprint{12600} test images of varying sizes, from $32\times32$ to $64\times64$ pixels, showing  traffic signs belonging to 43 different classes~\cite{stallkamp2011german}. We used  the MicronNet~\cite{wong2018micronnet} architecture as DNN.\\
\noindent\textbf{Computational Setup.}
We conducted the experiments using the deep learning library PyTorch~\cite{pytorch}. The experiments were conducted on a server with 4x NVIDIA A6000, an AMD EPYC 7773X CPU using 64 physical cores, and 768 GB of main memory.

\noindent\textbf{Model Architectures.}
Due to the absence of existing work on mitigating backdoor attacks on SL, we aligned the used DNN architectures on existing work about the security of Federated Learning~\cite{cao2021provably,fereidooni2024freqfed}. To ensure a comprehensive evaluation, we included 4 different model architectures in our evaluation with parameter sizes ranging from \numprint{440812} to \numprint{13425930} trainable parameters. The details of the used DNN architectures are shown in Tab.~\ref{tab:modelArchitectures}. We focused our evaluation on the \cifar dataset and evaluated all DNN architectures for this dataset (see \sect\ref{sec:eval-models}). In addition, we also evaluated the \mnist and \fmnist datasets. Due to the simplicity of these datasets, we used a simple CNN for these datasets, as defined by Cao \etal~\cite{cao2021provably} for their evaluation of backdoor attacks against Federated Learning. Notably, the structure of this simple CNN slightly varies for different datasets, as the images in \cifar have a dimension of $32\times32$ pixels while \mnist and \fmnist consist of images with the dimensions $28\times28$.

\begin{table}[tb]
    \centering
    \caption{Overview of used Deep Neural Network (DNN) Architectures.}
	\label{tab:modelArchitectures}
	\scaleTable{
	\begin{tabular}{l|rrrr|l}
            & \multicolumn{4}{c|}{Number of Parameters} &\\
		Model & Head & Backbone & Tail & Total & \STAB{Evaluated\\ Datasets}\\\hline
            ResNet18~\cite{he2016deep}& \numprint{9536} & \numprint{11166976} & \numprint{5130} & \numprint{11181642} & \cifar\\
Simple CNN~\cite{cao2021provably} & \numprint{1520} & \numprint{665562} & \numprint{5130} & \numprint{672212} & \cifar \\
Simple CNN~\cite{cao2021provably} & \numprint{520} & \numprint{435162} & \numprint{5130} & \numprint{440812} & MNIST, FMNIST \\
GoogLeNet~\cite{szegedy2015going}& \numprint{124736} & \numprint{5475168} & \numprint{10250} & \numprint{5610154} & \cifar \\
VGG11~\cite{simonyan2015very} & \numprint{1920} & \numprint{9224064} & \numprint{4199946} & \numprint{13425930} & \cifar \\
Wide-ResNet50~\cite{zagoruyko2016wide}& {\numprint{9536}} & {\numprint{66824704}} & {\numprint{204900}}& {\numprint{67039140}} & {\cifarHundred} \\
MicronNet~\cite{wong2018micronnet}& {\numprint{824}} & {\numprint{411192}} & {\numprint{3010}}& {\numprint{415026}} & {\gtsrb}
	\end{tabular}}
\end{table}

\noindent\textbf{Training Parameters.} Unless stated otherwise, we considered a system consisting of 10 clients, from which 2 are malicious and aim to inject a backdoor. We simulated non-IID data on the client side using the main-label strategy that is frequently used in other work on distributed learning~\cite{cao2021fltrust,fereidooni2024freqfed}. Here, each client gets randomly assigned a main label. While a certain fraction, indicated by the IID rate, is samples from all available samples, the remaining samples are chosen only from the assigned main label class.  An IID rate of 1.0 indicates a complete IID data distribution among clients. As the default value, we use an IID rate of 0.8. Only for \gtsrb and \cifarHundred we used an IID-rate of 1.0 as default value to achieve a decent MA, as the number of clients here is significantly lower than the number of labels.

\noindent\textbf{{Considered Backdoor Behavior.}} In the following, we evaluate \ournameGen effectiveness for different datasets and backdoors. Unless stated otherwise, we use for \cifar a semantic backdoor that misclassified cars in front of a striped background as birds and for all other data sets a backdoor being activated by a red rectangle or, in the case of grayscale data sets such as \mnist, a white rectangle.

\subsection{Effectiveness of \ourname}
\ourname undergoes extensive evaluation against different backdoor attacks for pixel and semantic triggers, as depicted in Table~\ref{tab:effectiveness}. As the table shows, \ourname reduced the BA in cases to a negligible value. Notably, for pixel trigger backdoors, the BA is often even for a benign model not exactly 0\%. This phenomenon occurs because the model misclassified some test samples, and the metric counts them in favor of the backdoor if the predicted label is equal to the backdoor target label {(see App.~\ref{app:missclassificationsBA} for details)}. {Furthermore, we conduct thorough evaluations on the \cifar dataset in various attack scenarios, which we describe in the following.}

\begin{table}[b]
    \centering
    \caption{Effectiveness of \ourname against different attacks for the respective dataset, in terms of Backdoor Accuracy (BA) and Main Task Accuracy (MA).}
	\label{tab:effectiveness}
	\scaleTable{
	\begin{tabular}{l|l|rr|rr}
		& \multirow{2}{*}{Attacks}& \multicolumn{2}{c|}{No Defense} & \multicolumn{2}{c}{\ourname} \\
		\cline{3-6}
		Dataset& & BA & MA  &  BA & MA \\\hline
		\multirow{2}{*}{\mbox{\cifar}} & Semantic Trigger & 59.3 & 66.6 & 0.0 & 62.7\\
		  & Pixel Trigger & 100.0 & 66.6 & 0.3 & 66.4\\
		 MNIST & Pixel Trigger & 86.2 & 98.7 & 0.0 & 98.8\\
		 FMNIST & Pixel Trigger & 79.8 & 83.0 & 3.4 & 84.6\\
   \cifarHundred & {Pixel Trigger} & {93.3} & {76.8} & {0.1} & {76.5}\\
   {\gtsrb} & {Pixel Trigger} & {30.0} & {58.0} & {0.6} & {63.7}\\
	\end{tabular}}
\end{table}

\subsection{Different Data Scenarios}
\begin{figure}[b]
	\centering{	
			\includegraphics[width=0.775\columnwidth]{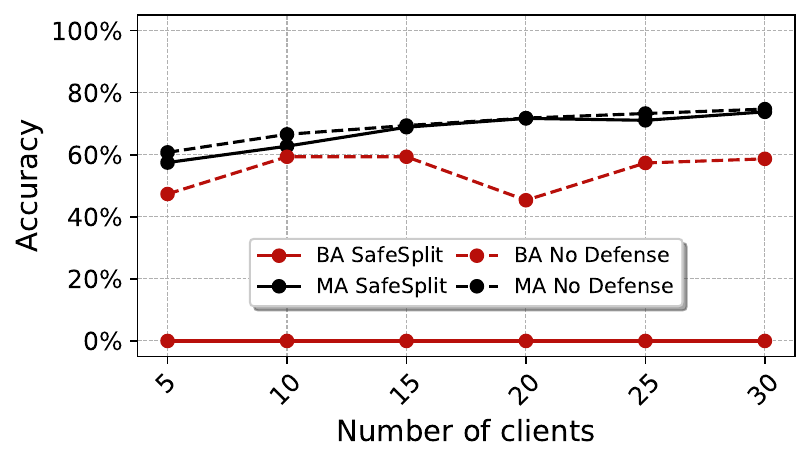}
	}
    
	\caption{BA and MA for different participant numbers.}
	\label{fig:eval-clientssize}
\end{figure}

We conducted different experiments on the \cifar dataset, varying the degree of non-IID data distribution to assess its effect on \ourname performance. The results are shown in Table~\ref{tab:iidRate}. Across IID rates of 0.6, 0.8, and 1.0, \ourname demonstrates significant reductions in BA while maintaining MA. Notably, for decreasing IID rates, the MA goes down regardless of the presence of an attack or defense, as training in such non-IID settings becomes very challenging.\\
\begin{table}[tb]
    \centering
    \caption{Effectiveness of \ourname for different degrees of IID data.}
	\label{tab:iidRate}
	\scaleTable{
	\begin{tabular}{r|rr|rr}
		& \multicolumn{2}{c|}{No Defense} & \multicolumn{2}{c}{\ourname} \\
		\cline{2-5}
		IID-Rate & BA & MA  &  BA & MA \\\hline
		0.6 & 44.7 & 59.5 & 0.0 & 57.5\\
		0.8 & 59.3 & 66.6 & 0.0 & 62.7\\
		1.0 & 60.7 & 69.4 & 0.0 & 68.1\\
	\end{tabular}}
\end{table}
Further, we assessed different numbers of clients being involved in the training process, ranging from 5 to 20, with 20\% of them being malicious. Fig.~\ref{fig:eval-clientssize} illustrates the BA and MA for \cifar. As the figure shows, \ourname effectively mitigates the attack, therefore keeping the BA 0\% and maintaining the MA with only a negligible drop compared to the scenario without a defense applied.

\subsection{Impact of Poisoned Model Rate (PMR)}
\begin{figure}[b]

	\centering{	
			\includegraphics[width=0.775\columnwidth]{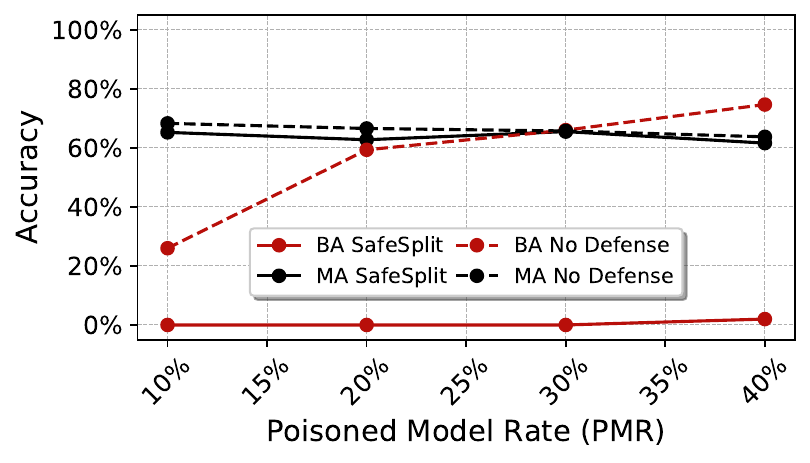}
	}
	\caption{BA and MA for various Poisoned Model Rates (PMRs).}
	\label{fig:eval-pmr}
\end{figure}

Fig.~\ref{fig:eval-pmr} shows \ournameGen effectiveness for varying ratios of malicious clients (PMR). As the figure shows, \ourname effectively mitigates the backdoor, keeping the BA at 0\%, while maintaining the MA close to the scenario without defense. Notably, the figure also shows that the attack is less effective for low PMRs.

\subsection{Adaptive Attacks}
\label{sec:eval-adap}
A sophisticated adversary that is aware of the defense might adapt the attack to make it more effective against the deployed defense approach. In the following section, we describe and evaluate various sophisticated attack strategies. Notably, some of the adaptive attacks assume a stronger adversary having knowledge of the current parameters of the backbone model, exceeding the previously defined threat model (see \sect\ref{sec:problem-advmodel}). In practice, such an adversary could, for instance, approximate the backbone’s parameters using a shadow model trained separately with the malicious clients’ benign data. However, for the sake of a comprehensive evaluation, the following section assumes that the adversary has knowledge of the actual backbone parameters.

\noindent\textbf{Varying the Poisoned Data Rate (PDR).}
To increase the similarity of the poisoned models to benign models, \adversary might vary the ratio of data samples for the backdoor behavior in its local dataset (Poisoned Data Rate, PDR). The choice of this parameter realizes a tradeoff since low PDRs ensure that the resulting model remains inconspicuous but also reduce the efficiency of the attack, while high values result in a high attack impact but also make the models easier to detect.
Tab.~\ref{tab:pdr} shows the effectiveness of \ourname for varying PDRs. As the table shows, \ourname effectively mitigates the attack for all PDR values. Notably, the table also shows that the attack is, {in the absence of a} defense, most effective for a PDR of 50\%, while for higher PDR values, the BA is reduced. Although this might be counterintuitive at first glance, a reason for this might be that for high PDRs, the dataset of the malicious clients becomes highly imbalanced and focuses on the single backdoor target label. This imbalanced data causes the model to overfit significantly and results in changes that can be easily reverted by the benign clients through their training.
\\
\noindent\textbf{Loss Constrain for Rotation Distance Metric.} A key aspect of \ournameGen backdoor detection is the rotational distance metric. Following existing attacks on Federated Learning~\cite{bagdasaryan}, a sophisticated adversary might adapt the loss function and integrate a regularization term into the loss function that minimizes the rotational distance of the current poisoned model to a reference model. Given the loss function $\mathcal{L}_{\text{class}}$ that measures the model's performance on its training data, the anomaly evasion loss $\mathcal{L}_{\text{ano}}$ that measures the suspiciousness of the model, here the rotational distance to the reference model, then according to Bagdasaryan \etal the combined loss function is defined as
\begin{equation}
    \mathcal{L} = \alpha \cdot \mathcal{L}_{\text{class}} + (1-\alpha)\cdot \mathcal{L}_{\text{ano}}
\end{equation}
where $\alpha$ is a hyperparameter that weights both terms. 

We evaluated this attack using the base model (\adversary has knowledge of the backbone) as reference model. Notably, knowing the backbone is not a realistic threat scenario, as the adversary would need to have access to the server, going beyond our threat model (see \sect\ref{sec:problem-advmodel}). Tab.~\ref{tab:alpha} shows the results for different $\alpha$-values. Although the adversary reduces the suspiciousness for the rotational distance, due to the combination of a static and dynamic analysis \ourname still effectively detected the poisoned models and reduced the BA to 0\%. Notably, setting $\alpha=0.5$ achieves in the absence of a defense the best BA without significantly degrading the MA. Thus, in the following experiments, we use $\alpha=0.5$.

\begin{table}[H]
    \centering
    \caption{{Effectiveness of \ourname against loss-constrain using different $\alpha$-values.}}

	\label{tab:alpha}
	\scaleTable{
	\begin{tabular}{l|rr|rr}
		& \multicolumn{2}{c|}{No Defense} & \multicolumn{2}{c}{\ourname} \\
		\cline{2-5}
		& BA & MA  &  BA & MA \\\hline
		$\alpha = 0.25$ & 62.0\% & 66.3\% & 0.0\% & 63.5\%\\
		$\alpha = 0.50$  &  60.0\% & 66.4\% & 0.0\% & 64.1\%\\
		$\alpha = 0.75$ & 51.3\% & 65.9\% & 0.0\% & 63.9\% \\
	\end{tabular}}
\end{table}

To evaluate also a more realistic setting, we repeated the experiment using a separate backbone as reference that was trained from the same base model. However, also here \ourname was able to detect the poisoned model updates, resulting in a BA of 0\%.

\noindent\textbf{Loss Constrain for Static Distance Metric.} Analogous to the previous paragraph we added a regularization term that focuses on the frequency analysis of \ourname as {regularization term}. Again, we evaluated the attack using the base model and also a benign model as reference models. In both cases, we extracted the low frequencies of the Discrete Cosine Transform of the reference model and the poisoned model under training; we then constrained the loss function to minimize the Euclidean Distance between the two frequency representations. However, due to the characteristics of the Discrete Cosine Transform domain, even small changes in the low-frequency components of the model can result in significant changes in the model's parameters. The regularization terms of the loss function, can either try to only inject the backdoor behavior (with $\alpha$ values close to 1) or try to bring the malicious model DCT as similar as possible to a benign one (with $\alpha$ values close to 0). Therefore, in the first case, the model behaves again as a malicious model, and it is detected by \ourname, and in the second case the adversary fails to implant any meaningful backdoor, with a BA of 0\%. A third case is a balance of the two regularization terms (with $\alpha$ values close to 0.5), but in this scenario, the changes to the low frequencies produce a scrambled model with very unnatural behavior. Therefore, as \ourname is highly sensitive to dynamic changes in model behavior, it was still able to detect the backdoored models and reduced the BA to 0\% in both cases.

\begin{table}[b]
    \centering
    \caption{Effectiveness of \ourname for different Poisoned Data Rates (PDRs).}

	\label{tab:pdr}
	\scaleTable{
	\begin{tabular}{r|rr|rr}
		& \multicolumn{2}{c|}{No Defense} & \multicolumn{2}{c}{\ourname} \\
		\cline{2-5}
		PDR-Rate & BA & MA  &  BA & MA \\\hline
		25\% & 85.3 & 68.4 & 0.0 & 64.7\\
		50\% & 88.7 & 67.9 & 0.0 & 65.1\\
		75\% & 59.3 & 66.6 & 0.0 & 62.7\\
		100\% & 24.0 & 51.0 & 0.0 & 64.5\\
	\end{tabular}}
\end{table}
\noindent\textbf{Loss Constrain for Static and Dynamic Distance Metrics.} Building on the two previous attacks, we integrated both loss functions as regularization terms. However, we observed \ourname to remain effective and reduce the BA to 0\% for both reference models (base model and trained benign model). This might be caused by the trade-off that the DNN optimizer needs to perform during the training. Because of the regularization term, the model must not show any indications of contradicting (backdoor) behavior. However, at the same time, due to the original loss that focuses on optimizing the predictions, the model is trained to show backdoor behavior. The optimizer then aims to minimize both aspects and perform a trade-off between both aspects. However, the resulting model will then still show poisoned behavior, allowing \ourname to detect it.

\noindent\textbf{Loss Constrain for Euclidean Distance.} An alternative option is to constrain the model not with respect to the specific metric that the defense uses but to integrate the Euclidean distance into the loss function to keep all parameters of the model close to the respective reference model. However, as this technique needs to perform a trade-off again, {it still needs to train the model on the backdoor behavior, \ourname detected the nuances that} indicate the poisoning and reduced the BA to 0\%.

\noindent\textbf{Focus Training on Tail.} \ourname analyses only the backbone model, as the head and tail are held by the clients, and the server has no access to them in order to preserve the client's privacy. However, a sophisticated adversary might try to exploit this and train the backdoor only into the tail. To achieve this, the malicious clients train the model alternating with batches containing only benign samples and samples also containing samples for the backdoor behavior. For the batches containing backdoor behavior, the client uses the server for forward propagation, thus making the predictions. However, after calculating the loss for the poisoned batches, the backpropagation and gradient descent are only applied to the tail, and the gradients are not shared with the server. As a result, the backbone and head will be trained only with clean data, while the tail is trained for benign and poisoned data. In this case, the backbone would be inconspicuous. However, as our experiments showed, for this attack strategy, the backdoor is not injected into the model, even without defense, and the BA remains 0\%.

\subsection{Comparison with Defenses for Federated Learning}\label{sec:eval-sota}
\begin{figure}[tb]
	\centering{
		\subfloat[Backdoor Accuracy]{		
			\includegraphics[width=0.775\columnwidth]{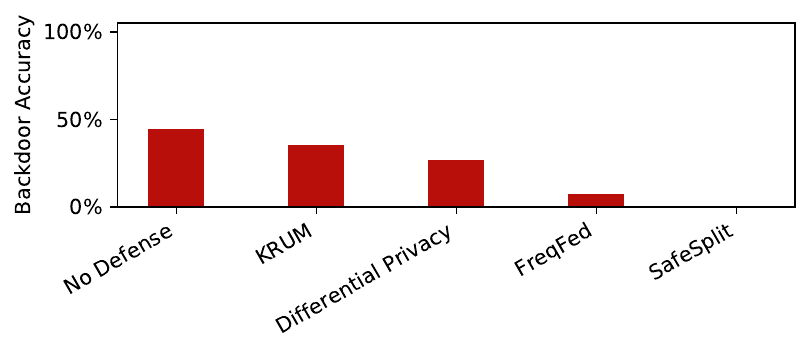}
			\label{fig:eval-cmpdef:ba}
		}\\
		\subfloat[Main Task Accuracy]{
			\includegraphics[width=0.775\columnwidth]{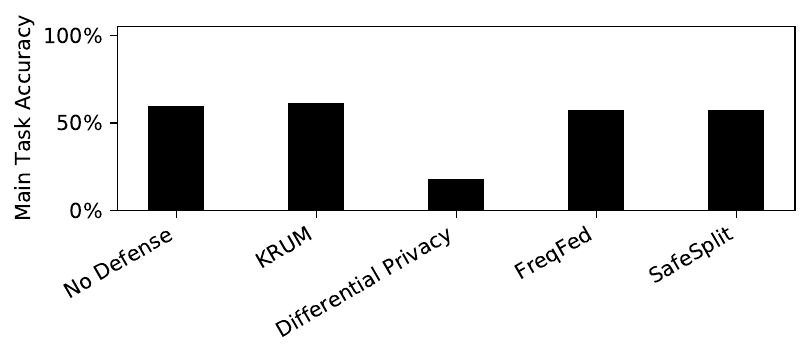}
			\label{fig:eval-cmpdef:ma}
		}
	}
	\caption{Comparison of different \sota defense techniques against \ourname using Main Task Accuracy (MA) and Backdoor Accuracy (BA).}
	\label{fig:eval-cmpdef}
\end{figure}

While the existing literature does not provide defenses against client-side backdoor attacks in SL, many defenses were proposed for the distributed learning scheme Federated Learning. Based on the categorization of Fereidooni \etal~\cite{fereidooni2024freqfed}, we selected 3 representative backdoor defenses for FL and adapted them for SL. Particularly, we evaluated FreqFed~\cite{fereidooni2024freqfed} as an approach that aims to classify all benign and poisoned models, KRUM~\cite{blanchard17Krum} that focuses on correctly identifying a subset of benign models, and Differential Privacy~\cite{bagdasaryan,mcmahan2018iclrClipping} that aims to mitigate the backdoor without identifying the attackers.
{Notably, defending backdoor attacks in SL raises significant challenges (see \sect\ref{sec:problem-challenges}), which \ourname addresses. To avoid any unjustified disadvantages for the existing techniques, we adapted them, using Euclidean distance of the updates for Krum and integrated \ournameGen circlewise structure into FreqFed. For Differential Privacy no adaptions were necessary, as noising and clipping are independent of the SL's structure. Fig.~\ref{fig:eval-cmpdef} shows the BA and MA values for a \nonIid scenario using an IID rate of 0.6. All defenses except Differential Privacy maintain a comparable MA. However, despite the adaptions, only \ourname is able to mitigate the attack and reduce the BA to 0\%.}

\subsection{Performance in the Absence of Attacks.}
An important objective for a practical defense is to not negatively affect the training process (see O2 in \sect\ref{sec:problem-challenges}). To measure the impact of \ourname on the training of a model, we conducted several experiments starting from a random base when no attack is performed and no defense is applied or \ourname is deployed. After 50 rounds of training without defense, the MA reached 69.30\%, while when \ourname was used, the MA achieved 66.6\%; thus, also, when deploying \ourname, no significant drop in MA was observed.

\subsection{Performance for Different Model Architectures.}
\label{sec:eval-models}
We evaluated the effectiveness of \ourname using different model architectures. Tab.~\ref{tab:modelsEffectiveness} shows the effectiveness of \ourname for 4 different model architectures that are frequently used in distributed learning~\cite{cao2021provably,fereidooni2024freqfed,gupta2018distributed}. As the table shows, \ourname successfully mitigates the backdoor for all models while maintaining the MA on a similar level as without the attack.

In addition, we also evaluated different positions for the cutting layer for the ResNet-18 model to simulate different backbone sizes. Resnet-18 consists of a convolutional layer a batch normalization layer, a ReLu layer, a max-pooling layer, 4 blocks and a linear layer. Each block consists of 4 convolutional layers, several batch normalization layers, and a down-sampling layer. We simulated different backbone sizes by assigning different numbers of blocks (2,3,4) to the backbone and clients. We observed \ourname to effectively mitigate the attack and reduce the BA to 0\%.\\

In summary, we evaluated \ourname for different datasets and data scenarios, attack types and settings, and client numbers and compared it against different baseline defenses. \ourname was always able to mitigate the backdoor attack and reduced in all experiments the BA to less than 5\%.

\section{Security Considerations}
\label{sec:discussion}
\noindent In the following, we discuss the security aspects of our scheme to fulfill our security and functional objectives and challenges (cf. \sect\ref{sec:problem-advmodel}). 

To address objective O1 (Prevent Backdoor Attacks), a backdoor defense must fulfill the security requirement of significantly reducing the attack impact. In the following, we will first discuss the risk of server-side attacks, before summarizing \ournameGen resilience against backdoor attacks (Backdoor Resilience), {including its robustness} against defense-adapted attack strategies that aim to make the DNNs' parameters inconspicuous (Adaptive Attacks) as well as attack strategies that aim to prevent that the actually poisoned parameters are analyzed by the defense (Analysis Evasion).\\

\noindent\textbf{Effectiveness against Different Adversary Models.}
We introduced the first defense against backdoor attacks in Split Learning, focusing on client-side attacks. As mentioned in Section~\ref{sec:problem-advmodel}, server-side backdoor attacks have already been the subject of investigation in the literature~\cite{erdogan2022splitguard}. In contrast, client-side backdoor attacks remained an open challenge, as we discussed in Section~\ref{sec:problem-advmodel}.  Moreover, in typical practical settings, we expect the server side to be a well-protected cloud instance, unlike the client side, which can be any mobile device with less protection. In \sect\ref{sec:eval}, we extensively evaluated \ourname in various scenarios using different attack settings. \\
\begin{table}[tb]
    \centering{
    \caption{Effectiveness of \ourname for different Deep Neural Network Architectures using the \cifar dataset.}
	\label{tab:modelsEffectiveness}
	\scaleTable{
	\begin{tabular}{l|rr|rr|rr}
		& \multicolumn{4}{c|}{No Defense} & \multicolumn{2}{c}{\ourname} \\
		& \multicolumn{2}{c|}{No Attack}& \multicolumn{2}{c|}{Attack} & \multicolumn{2}{c}{Attack} \\
		Model & BA & MA &  BA & MA &  BA & MA \\\hline
		ResNet-18 & 0.0 & 69.3 & 59.3 & 66.6 & 0.0 & 62.7\\
		Simple CNN & 0.0 & 64.1 & 78.0 & 62.7 & 0.0 & 60.4\\
		GoogLeNet & 0.0 & 63.5 & 16.7 & 57.9 & 0.0 & 60.2\\
		VGG11 & 0.0 & 47.6 & 76.7 & 49.7 & 0.0 & 43.0\\
	\end{tabular}}
    }
\end{table}

\noindent\textbf{Backdoor Resilience.} In the \sect\ref{sec:eval}, we evaluated the effectiveness of \ourname in various scenarios and attack settings. Particularly, we used five \sota image recognition benchmark datasets (\cifar, \mnist, \fmnist, \gtsrb, \cifarHundred; see Tab.~\ref{tab:effectiveness}), different numbers of clients (see Fig.~\ref{fig:eval-clientssize}), IID settings (see Tab.~\ref{tab:iidRate}), 4 different model architectures (see Tab.~\ref{tab:modelsEffectiveness}) and various scenarios. In addition, we evaluated different attack settings, particularly different ratios of poisoned data (see Tab.~\ref{tab:pdr}) and attack strategies (see \sect\ref{sec:eval-adap}). We observed that in all cases \ourname mitages the attack successfully, thus achieving O1 (see \sect\ref{sec:problem-advmodel}).

\noindent\textbf{Adaptive Attacks.}
A powerful adversarial evasion strategy concerns manipulating the training process to ensure the analyzed backbone model does not show high angular distance values. In this paper, we consider a sophisticated adversary, being aware of the defense and having full control over the individual clients (see \sect\ref{sec:problem-advmodel}). In \sect\ref{sec:eval-adap}, we evaluated several defense agnostic attacks that aim to circumvent \ourname by leveraging loss-constrain and integrating the angular distance and frequency distance into the loss function, restricting the distance to benign models, and training the backbone only with benign data. However, \ourname was able to successfully mitigate all of these attacks, showing \ournameGen robustness even against sophisticated adaptive attack strategies.\\
\noindent\textbf{Analysis Evasion.} {In our adversary model, the} adversary has full control over the head and tail (see \sect\ref{sec:problem-advmodel}) and can, before forwarding them to the next client, even entirely replace the values. Thus, the forwarded values are not necessarily those obtained by the training. \ourname counters this attack by analyzing the backbone stored on the server, where they can only be changed in a well-controlled manner (backpropagation). This ensures that the model used for the following clients is the same for backdoor detection, effectively preventing time-to-check-time-to-use attacks. Another strategy to distract angular metrics in other distributed settings is upscaling the model's parameters without changing them~\cite{bagdasaryan,bhagoji}. However, as only the server has access to the backbone, this strategy is restricted to the head and tail that reside at the clients. Therefore, the backbone would need to remain inconspicuous to circumvent \ourname. However, in \sect\ref{sec:eval}, we demonstrated that changing the backbone is essential for injecting the backdoor, such that \ourname is also robust against scaling attacks.
Thus, \ourname effectively and significantly reduces the risk of adversaries injecting backdoors into the model while marginally impacting the benign main task accuracy, fulfilling our functional and security objectives and requirements.

\section{Related Work}
\label{sec:sota}
\label{sec:related_works}
\noindent This section provides an overview of the recent research progress in the fields of privacy, security, and reliability for split learning.
\noindent\\
\textbf{Security of Split Learning.}
As mentioned in Sect.~\ref{sec:intro}, split learning has emerged as a promising {alternative to Federated Learning,} offering substantial reductions in the computational resources required by participants \cite{gao2020end, singh2019detailed, thapa2021advancements}. However, the inherent data and model control separation in split learning has raised various security concerns that can be categorized as the vulnerability to: data reconstruction~\cite{pasquini2021unleashing,erdogan2022splitguard,gao2023pcat,yu2024chronic}, label inference attacks~\cite{li2021label,fu2022label,liu2024similarity,bai2023villain}, and backdoor attacks~\cite{tajalli2023feasibility,yu2023backdoor,he2023backdoor,bai2023villain}. While we only focus on the backdoor attacks in this paper, we briefly explain the other attack vectors for completeness. 

In a data reconstruction attack, the adversary resides on the server and tries to recover the original training data from the feature vector uploaded by the clients. These attacks are further divided into two categories~\cite{yu2024chronic}: passive attacks, where the server does not disrupt the standard training process, only leveraging the intermediate steps of the training to gain information about the samples and create an attack model~\cite{erdogan2022splitguard,gao2023pcat}. Active data reconstruction attacks instead manipulate the training process to reconstruct the client data~\cite{pasquini2021unleashing,yu2024chronic} and obtain better results but faces the risk of being more easily detectable by the client defenses~\cite{erdogan2022splitguard,fu2023focusing}.  

In vanilla Split Learning, it is assumed that the clients cannot access the target labels. At the same time, the server cannot associate each label with specific samples to maintain privacy. Therefore, label inference attacks involve an adversary having control of {both, the} server and some clients to aim and steal label information of the data missing from the adversary's samples~\cite{li2021label,fu2022label,liu2024similarity,bai2023villain}. The defense mechanisms proposed to counteract label inference attacks primarily rely on random disturbance and differential privacy~\cite{yang2022differentially,liu2022clustering}.

Lastly, in the context of Split Learning, the objective of backdoor attacks shifts towards implanting backdoors into the connected model. However, the challenge differs depending on whether the adversary controls the server or the clients. In the first scenario, the server aims to successfully compromise the portion of the model that the clients possess and can attain this using surrogate clients~\cite{tajalli2023feasibility} or shadow models~\cite{yu2023backdoor}. In the client-side backdoor attack, the adversary aims to compromise {both, the} model residing in the server and, as an additional challenge, the backdoor needs to persist even on the portion of models possessed by the other victim clients. To achieve this, the malicious clients can employ auxiliary models~\cite{yu2023backdoor} or poison the training by submitting ad-hoc trigger vectors during training~\cite{he2023backdoor,bai2023villain}. However, we stress that the existing literature on client-side backdoor attacks was proposed only for the vanilla Split Learning framework, where it is assumed that the clients do not have access to ground truth label information~\cite{bai2023villain}. Hence, mitigating client-side attacks in SL has remained an open challenge we aim to tackle in this paper for the first time, especially in the context of U-shaped SL. 

\noindent\\
\textbf{Backdoor Defenses in Federated Learning (FL).} In the context of other distributed collaborative learning paradigms, such as FL the objective of backdoor attacks is to implant backdoors into the global model. However, the challenge lies in preserving the efficacy of the local model's backdoor post-aggregation on the server side. Notable contributions in this domain include the works of Bagdasaryan \etal~\cite{bagdasaryan} and Xie \etal~\cite{dbabackdoor}, Saha \etal~\cite{saha2020hidden}, Shumailov \etal~\cite{shumailov2021manipulating}, and Wenger \etal~\cite{wenger2021backdoor}, aim to compromise model integrity through subtle or overt manipulation of the training data. 

Multiple defenses have been proposed to mitigate these attacks in {FL}~\cite{fereidooni2024freqfed,fung2020FoolsGold,blanchard17Krum,bagdasaryan, wang2022flare,cao2021provably,li2023flairs}. KRUM calculates the pairwise Euclidean Distances and, analogously to \ourname, sums up the neighborhood to obtain a score. Afterward, the model with the lowest KRUM score is selected as the aggregated model. However, the Euclidean distance can be circumvented as shown by various works for {FL}~\cite{fereidooni2024freqfed} and in \sect\ref{sec:eval} for SL. Further, in SL, it is important to select the latest model rather than the model with the smallest score. FreqFed~\cite{fereidooni2024freqfed} also leverages the observation that training a model for new behavior results in large changes in a model's low-frequency components. To detect backdoored models, clustering is applied to the models' low frequencies. However, FreqFed is not applicable to the Split Learning domain due to the sequentiality of the training steps and other challenges as mentioned in \sect\ref{sec:problem-challenges}. Bagdasaryan \etal~\cite{bagdasaryan} consider using Differential Privacy, but as our evaluation \sect\ref{sec:eval} shows, neither adding random noise to the backbone nor restricting the \lnorm of the updates helps in counteracting an advanced attack in SL. FoolsGold~\cite{fung2020FoolsGold} is a {FL}defense that assumes high similarity between model updates submitted by the adversary, penalizing clients that submit similar (sybils) updates during aggregation. As we discussed in \sect\ref{sec:intro} differently from {FL}, each client does not start training from the same global model, but instead, each starts training from a different base model. Therefore, malicious clients will each impact in different ways the updates on the backbone. Wang \etal~\cite{wang2022flare} propose a model poisoning defense that analyzes the latent space of the second to last layer of the DNN, to detect malicious behavior. However, as we describe in \sect\ref{sec:problem-system} in the U-shaped SL framework, the server does not have access to the last layers, which are instead controlled by the clients, further as outlined in \sect\ref{sec:problem-challenges} the server should not have access to any training or validation data. Therefore, it would be unable to analyze the behavior of the last layers due to its lack of access to data. Cao \etal introduced a secure aggregation protocol, where each client is assigned into random subsets of clients, and for each subset a separate distributed learning process is executed, resulting in separate models for each subset. Then, during inference, each global model is queried separately, and the final label prediction is determined by majority voting. Unfortunately, repeating the training process and inference for each sample multiple times results in an impractical overhead. Furthermore, the defense has been broken for a non-tiny number of malicious clients. Furthermore, in the case of highly \nonIid data or a limited number of participant clients, dividing the training into multiple subsets will impact the overall MA of the trained models.\\
In comparison, \ourname employs different metrics to detect backdoored models by analyzing models from the static and dynamic perspectives. The circlewise rollback mechanism allows skipping detected poisoned models while choosing the latest benign model to prevent reverting benign training contributions. 

\section{Conclusion}
\label{sec:con}
In this paper, we proposed \ourname, the first defense against client-side backdoor attacks in SL. Unlike existing methods for other distributed learning schemes, \ourname employs a rollback mechanism in which we conduct static (frequency) and dynamic (rotational) analysis to detect poisoned model updates and limit attack impact. Our extensive evaluation demonstrated \ourname's effectiveness across various scenarios, attack settings, and defense-agnostic strategies.


\section*{Acknowledgment}
This research received funding from the Horizon program of the European Union under grant agreements No. 101093126 (ACES) and No. 101070537 (CROSSCON), OpenS3 lab, as well as the Federal Ministry of Education and Research of Germany (BMBF) within the IoTGuard project and Athene projects.



%
\bibliographystyle{plain}
\bibliography{main}

\appendix
\subsection{Deep Neural Network (DNN)}
\label{app:dnn}
A DNN is a mathematical function denoted as $F(X; W)$, with $X$ representing the input data samples and $W$ denoting the network's parameters (comprising weights and biases). This network is structured into several layers, denoted as $F_i$, where $i \in \{1, \cdots, L\}$. The first layer, the input layer, is labeled as $F_1$, while the final layer, termed the output layer, is designated as $F_L$. The intermediate layers are commonly referred to as hidden layers. In a feed-forward neural network, data moves in a unidirectional path, starting from the input layer, traversing through the hidden layers, and ultimately reaching the output layer.

\subsection{Poisoning Attack}
\label{app:background-poisoning}
In machine learning security, poisoning attacks represent a scheme in which model parameters are intentionally manipulated during or after training to introduce abnormal behaviors. Targeted or backdoor attacks alter the DNN's training stealthily to generate specific mispredictions when the model is presented with inputs containing predetermined triggers. For example, a trigger could be a red pixel placed in the upper left corner of an input image, which must be incorporated into the training dataset through data poisoning, and the abnormal behavior would be the classification of all samples with the red pixel into a specific predetermined class. The success rate for the backdoor attack is calculated based on the prediction performance on triggered data, denoted as Backdoor Accuracy (BA), while maintaining the expected behavior on benign inputs, as indicated by a high model or main-task accuracy (MA).

Following Fig.~\ref{fig:slSplit}, we can see how an adversary must manipulate one or more clients to execute backdoor attacks in the Split Learning configuration. The adversary trains the poisoned local head and tail portion to poison the backbone in the central server and the subsequent head and tail training of the future clients. The critical objective is to poison the U-shaped aggregated model's prediction without incurring any noticeable behavior change to the server backbone and the other clients' head and tail, thus undermining the integrity and reliability of the entire system~\cite{he2023backdoor,bai2023villain}.

\subsection{Discrete Cosine Transform}
\label{app:background-dct}
In signal processing, the Discrete Cosine Transform (DCT)~\cite{Ahmed1974} is employed to break down a signal into its frequency components, providing insights into the underlying dynamics and transitions within the signal\cite{Shu2017}. The DCT takes a sequence of numbers and represents them as a combination of simple wave patterns (sinusoids) with different frequencies and sizes.

Mathematically, DCT transformations are invertible functions that convert an input sequence of $N$ real numbers into the coefficients (values that multiply the wave patterns) of $N$ orthogonal cosine basis functions (wave patterns independent of each other) with increasing frequencies. The DCT components (the coefficients) are ordered by significance (importance), with the first coefficient representing the sum of the input sequence normalized by length. Lower-order coefficients correspond to lower signal frequencies (slower repeating waves) and indicate the sequence's patterns (general trends or shapes in the data). These are known as low-frequency components. The 2-D DCT of a signal $S$ (e.g., a matrix of size $N$ by $M$) at frequencies $k$ and $l$ ($X(k,l)$), is given by the following equation~\cite{Chen1977DCT, Wang1984DCT, ahmed1974discrete,fereidooni2024freqfed}.

{\footnotesize
\begin{equation}
    X(k,l)=a_{k}a_{l}\sum_{m=0}^{M-1}\sum_{n=0}^{N-1}S(m,n) cos\biggl(\frac{k\pi }{2M}(2m+1)\biggr) cos\biggl(\frac{l\pi }{2N}(2n+1)\biggr)
\end{equation} 
}
\begin{figure}[tb]
	\centering{
		\subfloat[Rotational Distance Scores]{		
			\includegraphics[width=0.8\columnwidth]{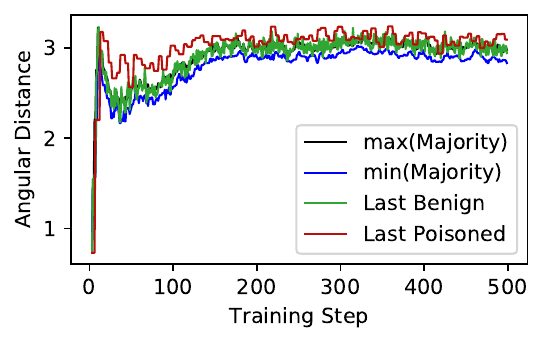}
			\label{fig:eval-scores-iid1:rot}
		}\\
		\subfloat[Frequency Distance Scores]{
			\includegraphics[width=0.8\columnwidth]{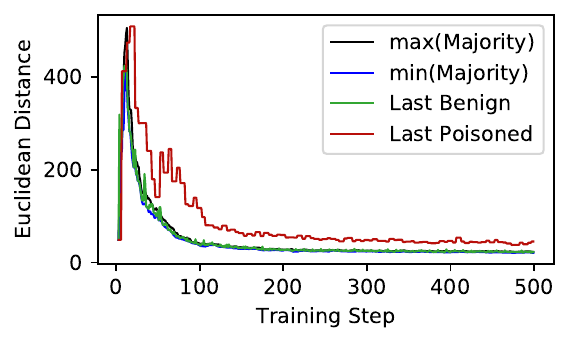}
			\label{fig:eval-scores-iid1:feq}
            }
	}
	\caption{Rotational and Frequency distance scores for an IID-rate of 1.0.}
	\label{fig:eval-scoresIID}
\end{figure}
\noindent \hspace{-0.1cm}Where $a_{k}, a_{l}$ are dependent on the values of $k$, and $l$, with the following rules:

{\footnotesize
\begin{equation}
a_{k} = \begin{cases}
\sqrt{\frac{2}{MN}} &\text{for $k=0$} \\
1 & \text{for $k=1,2,...,M-1$}
\end{cases}
\end{equation}}

{\footnotesize
\begin{equation}
a_{l} = \begin{cases}
\sqrt{\frac{2}{MN}} &\text{for $l=0$} \\
1 & \text{for $l=1,2,...,N-1$}
\end{cases}
\end{equation}}

As we elaborate in \sect\ref{sec:approach-frequency}, recent work showed that in the early stages of the training, mostly the low-frequency components change while the high-frequency components change during the fine-tuning, when the model is already close to convergence~\cite{Rahaman2019spectralbias,xu2019training}. In \sect\ref{sec:approach-frequency}, we describe how this can be used to help detect models with injected backdoor behavior.

\subsection{Effectiveness of the Angular Distance}
\label{app:distanceScores}

\begin{figure}[tb]
	\centering{
		\subfloat[Rotational Distance Scores]{		
			\includegraphics[width=0.8\columnwidth]{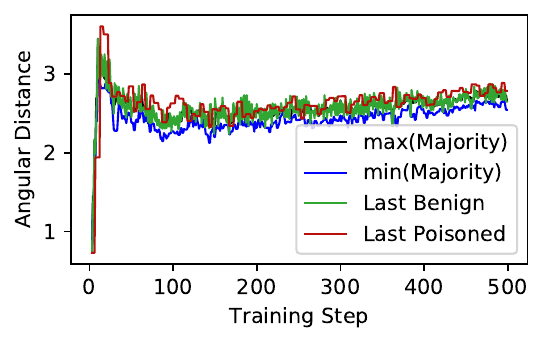}
			\label{fig:eval-scores-noniid:rot}
		}\\
		\subfloat[Frequency Distance Scores]{
			\includegraphics[width=0.8\columnwidth]{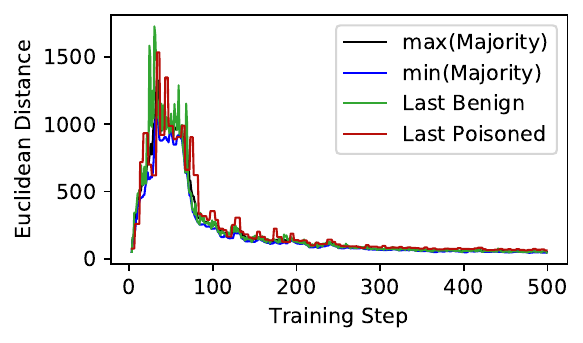}
			\label{fig:eval-scores-noniid:feq}
            }
	}
	\caption{Rotational and Frequency distance scores for an IID-rate of 0.8.}
	\label{fig:eval-scores-noniid}
\end{figure}

\ournameGen selection mechanism relies on the Euclidean distance in the frequency domain and the rotational distance among server model states. \ourname uses the assumption that the majority of clients are benign. We establish a norm by accepting a set of $\nicefrac{\nClients}{2}+1$ clients with the least {Euclidean} and rotational distances. Fig.~\ref{fig:eval-scoresIID} illustrates a distinct gap between malicious and benign scores for an IID-rate of 1.0 and Fig.~\ref{fig:eval-scores-noniid}. The figures visualize how the different perspectives (static and dynamic) complete it each other. As long as malicious scores exceed the maximum score of the majority set for at least one of both metrics, a malicious client remains unselected, preventing any backdoor injection. Only occasionally, a benign score may surpass the maximum of the majority set, resulting in its exclusion. This exclusion can impact MA, albeit minimally, as demonstrated in Table~\ref{tab:effectiveness}. Our evaluation reveals that the score gap between benign and malicious instances is more pronounced during the model's learning phase, gradually diminishing as MA converges. Hence, halting training at MA convergence proves advantageous. 

\subsection{Impact of Misclassifications on BA}
\label{app:missclassificationsBA}

For pixel-trigger backdoors the BA is often not 0\%, even when the model is benign and not poisoned. As outlined by Fereidooni \etal, this phenomenon occurs due to the misclassification of samples by the model, if the MA is not 100\%. Especially, samples that are similar to the backdoor target label are, if not recognized correctly, likely to be incorrectly classified as the backdoor target, even when they are independent of the presence of the backdoor trigger. For example, in the case of truck images, the model can misclassify them as car images. If a triggered sample is misclassified as the backdoor target by coincidence, it is counted as successful backdoor activation and increases the BA, although the model was trained to recognize the backdoor trigger~\cite{fereidooni2024freqfed}.

\begin{figure}[H]

	\centering{	
			\includegraphics[width=0.775\columnwidth]{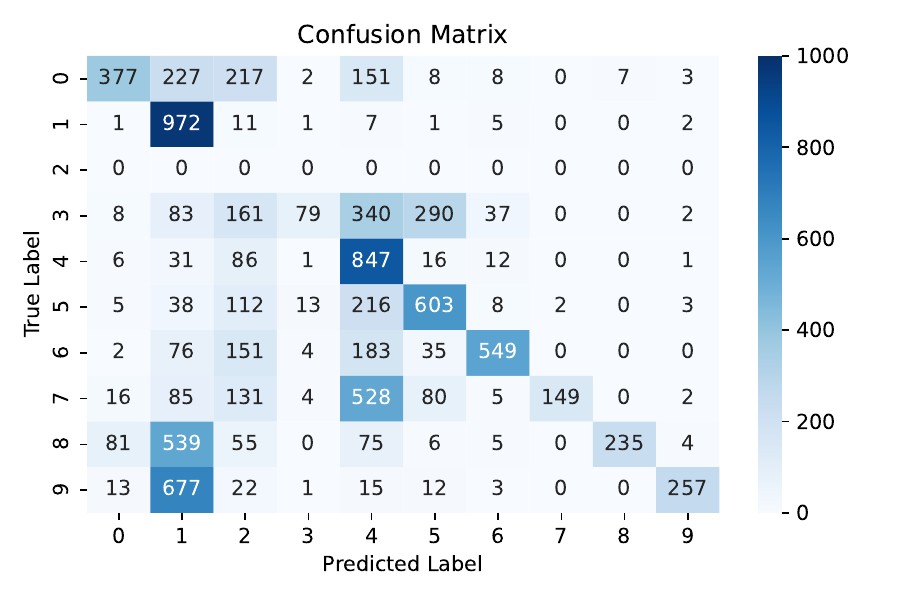}
	}
	\caption{Confusion matrix for triggered inputs applied on a benign model.}
	\label{fig:confusion}
\end{figure}

To illustrate this, we show in Fig.~\ref{fig:confusion} a confusion matrix for a poisoned test dataset evaluated on a benign model. In this case, the backdoor trigger, represented by a red rectangle, is intended to cause the model to predict class 2 (bird).

As the figure shows, although the model does not contain any backdoor, in 946 cases, label 2 is predicted, resulting in a BA of 10.5\%.

Notably, the red rectangle also covers parts of the image, thus affecting the model's ability to recognize the actual object.

\subsection{Evaluation of Further Attacks}
\label{app:further}
In addition, we evaluated \ourname for a label swapping attack, where the predictions for all samples of two classes shall be swapped, thus realizing a mixture of backdoor and untargeted attack. For this experiment, we measured the effectiveness using the attack success rate (ASR), counting the fraction of samples belonging to both classes where the model predicts the swapped labels. We performed this experiment for all pairs of classes using the \cifar dataset. We observed that \ourname effectively reduces the ASR in average from 22.7\% to 3.8\%. Notably, the ASR for the undefended model (22.7\%) is significantly smaller than for regular backdoor attacks such as the semantic backdoor. This is caused by the clean samples that are part of the benign clients' datasets, allowing them to reduce the attack's impact, while for backdoors, such as the semantic backdoor, benign clients pose only a negligible number of samples. Further, similar to the pixel-pattern backdoor, the BA for a benign model is not exactly 0\% but 3.4\% (cf. \sect\ref{app:missclassificationsBA}). 
While also untargeted attacks fall outside our threat model’s scope, since they can be straightforwardly detected and inherently differ from backdoor attacks, we tested a loss-maximization attack and showed that \ourname mitigates the attack and maintains the Main Task Accuracy (MA) at 64\%.

\subsection{Runtime Evaluation}\label{app:runtime}
\begin{figure}[H]

	\centering{	
			\includegraphics[width=0.775\columnwidth]{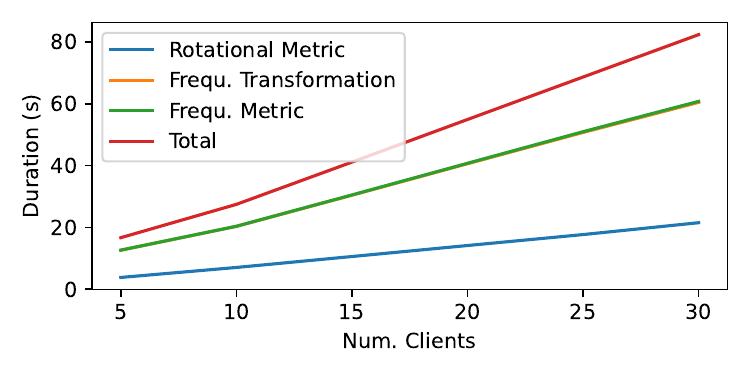}
	}
	\caption{Evaluation times of \ourname and its individual components for different client numbers.}
	\label{fig:evaluationTimes}
\end{figure}

We evaluated the runtime performance of \ourname and its individual components in dependence of the client number to determine its scalability. Since the defense is executed before every training, we performed for each client number an experiment running 50 rounds and measured the individual runtimes every time the defense was called. Thus, depending on the client number, we obtained measurements of 245 (5 clients) and 1470 (30 clients). Notably, we omitted measurements from the first round until every client provided at least one model. We trimmed the 5\% highest and smallest values and averaged the remaining values. The results are shown in Fig.~\ref{fig:evaluationTimes}. As the figure shows, the runtime of \ourname scales linearly as the most time-consuming operations, such as the frequency transformation, are executed once per model. Notably, the runtime for the frequency transformation is plotted as an individual curve but also included in the curve for the Frequency Analysis. 

Notably, all code was written in Python and no parallelization was used, although operations such as the frequency transformation can be easily parallelized. While runtime engineering is out of the scope of this work, this demonstrates the practical applicability of \ourname.

\subsection{Details on Rotational Distance}
\label{app:rotational}
The rotational distance metric is used to analyze the directional changes in the backbone’s parameter space over training rounds. It provides a way to measure how the configuration of the model’s parameters evolves dynamically over time. Unlike traditional magnitude-based metrics, which focus only on the size of updates, rotational distance captures the trajectory and orientation shifts of model updates, making it more robust against adversarial manipulations. 

\noindent
\textbf{Parameter Representation and Transformation: }The backbone of the model consists of high-dimensional weight tensors. Directly analyzing these tensors would be inefficient and difficult to interpret in terms of rotation. Instead, the rotational metric first extracts the backbone parameters and reshapes them into a structured representation that allows for spatial and directional analysis. This is done by first transforming the weight tensors into a 2D matrix form and then computing coordinate-based transformations (i.e., mapping into separate $x$- and $y$-coordinate vectors).

To create a coordinate space for the weights, we compute mean values along the rows and columns of the transformed weight tensors to obtain one-dimensional vectors of the mean values of rows and columns. Next, these obtained vectors are multiplied with the 2D matrix to preserve the variability of the values in the original backbone tensor. Finally, the obtained two 2D matrices are flattened to construct the $x$- and $y$-coordinate vectors {$B_t^x$ and $B_t^y$. This transformation allows the weight tensors to be entangled into x and y coordinate tensors, which will be used for computing angular displacement, as detailed in the following.

\noindent\textbf{Angular Displacement Computation:} Once the weight parameters are mapped into the coordinate space, the next step is to measure their angular displacement. The angular displacement $\theta(t)$ is defined as the coordinate-wise rotation of each paired $x$- and $y$ coordinate, obtained from $B_t^x$ and $B_t^y$ (i.e. rotation from the $x$-axis). This is computed as:  

\[
\theta(t) = \arctan \left( B_t^x, B_t^y \right)
\]

This formulation ensures that small shifts in weight direction are captured, even if the magnitude of the update remains the same. We apply a geometric transformation to the mapped weight tensors and compute the pairwise angular difference between successive updates. The use of $\arctan$ in this context is more than just a cosine similarity measure. It leverages the transformed $x$- and $y$-coordinate vectors to compute the angle between two vectors in a way that accurately captures the global directionality of parameter updates.

$\arctan$ is used to compute the orientation of the $x$- and $y$-coordinate vectors relative to the origin. This approach provides an equivalent measure of angular displacement but is expressed in direct coordinate-wise rotation (from the $x$-axis) rather than the computation of backbone similarity by utilizing the final $x$- and $y$-coordinate vectors. $\arctan$ determines the relative rotation of each coordinate point, effectively capturing local changes in parameter orientation.}

\noindent
\textbf{Estimating Rotational Frequency:} After computing the angular displacement, the rotational metric needs to determine how frequently these shifts occur over time. This is done by measuring the rate of angular displacement per unit time, which corresponds to angular velocity:

\[
\omega(t) = \frac{\theta(t) - \theta(t-1)}{\Delta t}
\]

where $\Delta t$ represents the time interval between consecutive updates. It is set to 1 in our approach. This provides an estimate of how quickly the orientation of the model parameters is changing. Finally, we convert the obtained angular velocity into rotational frequency, normalized by the full rotation cycle:

\[
RD = \frac{\omega(t)}{2\pi}
\]

This transformation ensures that the rotational distance is expressed in a form that captures repetitive shifts in parameter orientation over multiple training rounds. 

\noindent
\textbf{Pairwise Comparison of Rotational Frequencies: }Next, we compare the rotational frequencies across different models. The rotational distance metric \RDM is designed to track deviations from expected training trajectories by analyzing how different clients’ backbones behave over training rounds. Instead of comparing updates for a single model over time, it measures differences in rotational frequencies across multiple clients. The final rotational distance score {$R_i$ is obtained by summing the absolute differences:

\[
R_i = \sum | RD_i - RD_j |
\]}

Hence, \RDM allows the detection of anomalous client behavior, as adversarially manipulated models tend to exhibit higher frequency deviations compared to benign training updates.

\FloatBarrier
\vfill

\end{document}